\documentclass[prx,nosuperscriptaddress,twocolumn,longbibliography]{revtex4-1}

\usepackage{graphicx}
\usepackage{bm}
\usepackage{amssymb,amsmath,amsfonts,latexsym,dsfont}
\usepackage{upgreek}
\usepackage[usenames,dvipsnames]{color} 
\usepackage{stmaryrd}
\usepackage[english]{babel}
\usepackage{times}
\usepackage{appendix}

\usepackage[colorlinks=true , citecolor=blue,urlcolor=blue]{hyperref}

\usepackage[margin=1.5cm]{geometry}

\renewcommand{\emph}{\textit}

\newcommand{\id}{\mathds{1}}

\newcommand{\xa}{\alpha}

\newcommand{\xd}{\delta}
\newcommand{\xe}{\epsilon}

\newcommand{\xg}{\gamma}
\newcommand{\xt}{\theta}

\newcommand{\xr}{\rho}

\newcommand{\xph}{\phi}

\newcommand{\pp}{\perp}
\newcommand{\app}{\approx}

\newcommand{\Cs}{{}^{13}\R{C}}
\newcommand{\Ns}{{}^{14}\R{N}}

\newcommand{\bs}[0]{\mathbf \sigma}

\newcommand{\bn}[0]{\hat{\mathbf n}}

\newcommand{\mc}[1]{\mathcal{#1}}

\newcommand{\xD}{\Delta}

\newcommand{\xO}{\Omega}

\newcommand{\mU}[0]{\mathcal{U}}

\newcommand{\bx}[0]{\hat{\mathbf x}}

\newcommand{\bz}[0]{\hat{\mathbf z}}

\newcommand{\fr}[2]{\frac{#1}{#2}}
\newcommand{\ov}[1]{\overline{#1}}

\newcommand{\sq}[1]{\sqrt{#1}}

\newcommand{\rt}{\rightarrow}

\newcommand{\dg}{\dagger}

\newcommand{\beq}{\begin{equation}}
\newcommand{\eeq}{\end{equation}}
                  
\newcommand{\benum}{\begin{enumerate}}
\newcommand{\eenum}{\end{enumerate}}
                    
\newcommand{\bit}{\begin{itemize}}
\newcommand{\eit}{\end{itemize}}

\newcommand{\bea}{\begin{eqnarray}}
\newcommand{\eea}{\end{eqnarray}}

\newcommand{\bev}{\begin{verbatim}}
\newcommand{\eev}{\end{verbatim}}

\newcommand{\zt}{\times}

\newcommand{\qt}{\tau}

\newcommand{\lb}{\left(}
\newcommand{\rb}{\right)}
\newcommand{\lsb}{\left[}
\newcommand{\rsb}{\right]}
\newcommand{\lcb}{\left\{}
\newcommand{\rcb}{\right\}}

\newcommand{\pll}{\parallel}

\newcommand{\T}[1]{\textbf{#1}}
\newcommand{\I}[1]{\textit{#1}}
\newcommand{\R}[1]{\textrm{#1}}
\newcommand{\vv}{\vec}

\newcommand{\zl}[1]{\label{eqn:#1}}
\newcommand{\zr}[1]{Eq. (\ref{eqn:#1})}
\newcommand{\zfl}[1]{\protect\label{fig:#1}}
\newcommand{\zfr}[1]{Fig. \ref{fig:#1}}

\newcommand{\zar}[1]{Appendix \ref{app:#1}}

\newcommand{\ket}[1]{\left\vert{#1}\right\rangle}
\newcommand{\bra}[1]{\left\langle{#1}\right\vert}

\newcommand{\ba}{\left\{ \begin{array}{lr}}
\newcommand{\ea}{\end{array}\right.}

\newcommand{\Tr}[1]{\textrm{Tr}\left\{{#1}\right\}}

\newcommand{\blist}[1]{
 \begin{list}{#1}
 \begin{align}
	 arrow
 \end{align}
 $\checkmark\star
  { \setlength{\itemsep}{3pt}
     \setlength{\parsep}{2pt}
     \setlength{\topsep}{3pt}
     \setlength{\partopsep}{0pt}
     \setlength{\leftmargin}{1em}
     \setlength{\labelwidth}{1em}
     \setlength{\labelsep}{0.5em} } }
\newcommand{\elist}{
  \end{list}  }

\DeclareMathSymbol{\vartheta}{\mathalpha}{letters}{"12}
\DeclareMathSymbol{\theta}{\mathalpha}{letters}{"23}
\DeclareMathSymbol{\phi}{\mathalpha}{letters}{"27}
\DeclareMathSymbol{\varphi}{\mathalpha}{letters}{"1E}

\newcommand{\bef}
{
\begin{figure}[htbp]
\centering
}

\newcommand{\eef}{\end{figure}}

\newcommand{\beginsupplement}{%
        \setcounter{table}{0}
        \renewcommand{\thetable}{S\arabic{table}}%
        \setcounter{figure}{0}
        \renewcommand{\thefigure}{S\arabic{figure}}%
     }

\newcommand{\papertitle}{DC Magnetometry at the $T_2$ Limit}

\begin{document}


\title{\large{\papertitle}}

\author{A. Ajoy, Y. X. Liu and P. Cappellaro}
\affiliation{Research Laboratory of Electronics and Department of Nuclear Science \& Engineering,
 Massachusetts Institute of Technology, Cambridge, MA}

\begin{abstract}

Sensing static or slowly varying magnetic fields with high sensitivity and spatial resolution is critical to many applications in fundamental physics, bioimaging and materials science. Several versatile magnetometry platforms have emerged over the past decade, such as electronic spins associated with Nitrogen Vacancy (NV) centers in diamond.
However, their high sensitivity to external fields also makes them poor sensors of DC fields. Indeed, the usual method of Ramsey magnetometry leaves them prone to environmental noise, limiting the allowable interrogation time to the short dephasing time $T_2^*$. Here we introduce a hybridized magnetometery platform, consisting of a sensor and ancilla, that allows sensing static magnetic fields with interrogation times up to the much longer $T_2$ coherence time, allowing significant potential gains in field sensitivity. While more generally applicable, we demonstrate the method for an electronic NV sensor and a nuclear ancilla. It relies on frequency upconversion of transverse DC fields through the ancilla, allowing quantum lock-in detection with low-frequency noise rejection. In our experiments, we demonstrate sensitivities $\app 6\mu$T/$\sq{\textrm{Hz}}$, comparable to the Ramsey method, and narrow-band signal noise filtering better than $64$kHz. With technical optimization, we expect more than an one order of magnitude improvement in each of these parameters. Since our method measures transverse fields, in combination with the Ramsey detection of longitudinal fields, {it ushers in a compelling technique} for sensitive vector DC magnetometry at the nanoscale.

\end{abstract}
\maketitle

\section{Introduction}

Over the past two decades magnetic field sensors based on quantum systems have come of age, and  fields  at the few femto-Tesla level are now well within  detection reach. 
These exquisite sensitivities  often  result from exploiting quantum interference, as exemplified in a wide array of technologies, including SQUIDs~\cite{Clarke04}, atomic vapor cells~\cite{Kominis03,Budker07}, and Nitrogen Vacancy  (NV) center spins in diamond~\cite{Maze08,Balasubramanian08,Wolf15}. 
Spin qubits in the solid-state have shown their 
most sensitive performance  when the signals of interest are oscillating at frequencies of hundreds of kHz or higher (AC fields), 
due to the ability to operate the sensor away from intrinsic $1/f$ noise~\cite{Keshner82} that otherwise  plagues sensor performance. However, {arguably} many more signals of fundamental importance have low frequency  or are even static (DC fields). This includes  magnetic fields from biological processes, such as the firing of action potentials in single neurons~\cite{Wikswo80,Barry16,Jensen16}, and from magnetic materials, like edge currents in topological insulators~\cite{Wang15,Dellabetta12}. 
This is also true for a broader class of signals that can be transduced to effective magnetic fields, such as rotations~\cite{Ajoy12g}, pressure~\cite{Cai14} and electric fields~\cite{Dolde11}.

The conventional method for sensing DC magnetic fields is by the celebrated Ramsey technique~\cite{ramsey90}, where the sensor evolves freely acquiring a phase under the field to be measured. Unfortunately, the presence of noise  limits the sensor interrogation time (hence the total phase) to the {\it dephasing} time $T_2^{\ast}$. This is far shorter than the intrinsic sensor {\it coherence} time $T_2$, often by a few orders of magnitude~\cite{Wyk97}. 
Indeed, the key sensitivity advantage of AC sensors  derives  from lengthening the interrogation times up to  $T_2$, by performing phase sensitive (lock-in) detection~\cite{Scofield94,hill15} at the known signal frequency of interest, 
via a narrow bandwidth filter created by dynamical-decoupling (DD) sequences~\cite{Biercuk11, Ajoy11}.

\begin{figure}[h!]
  \centering
  {\includegraphics[width=0.49\textwidth]{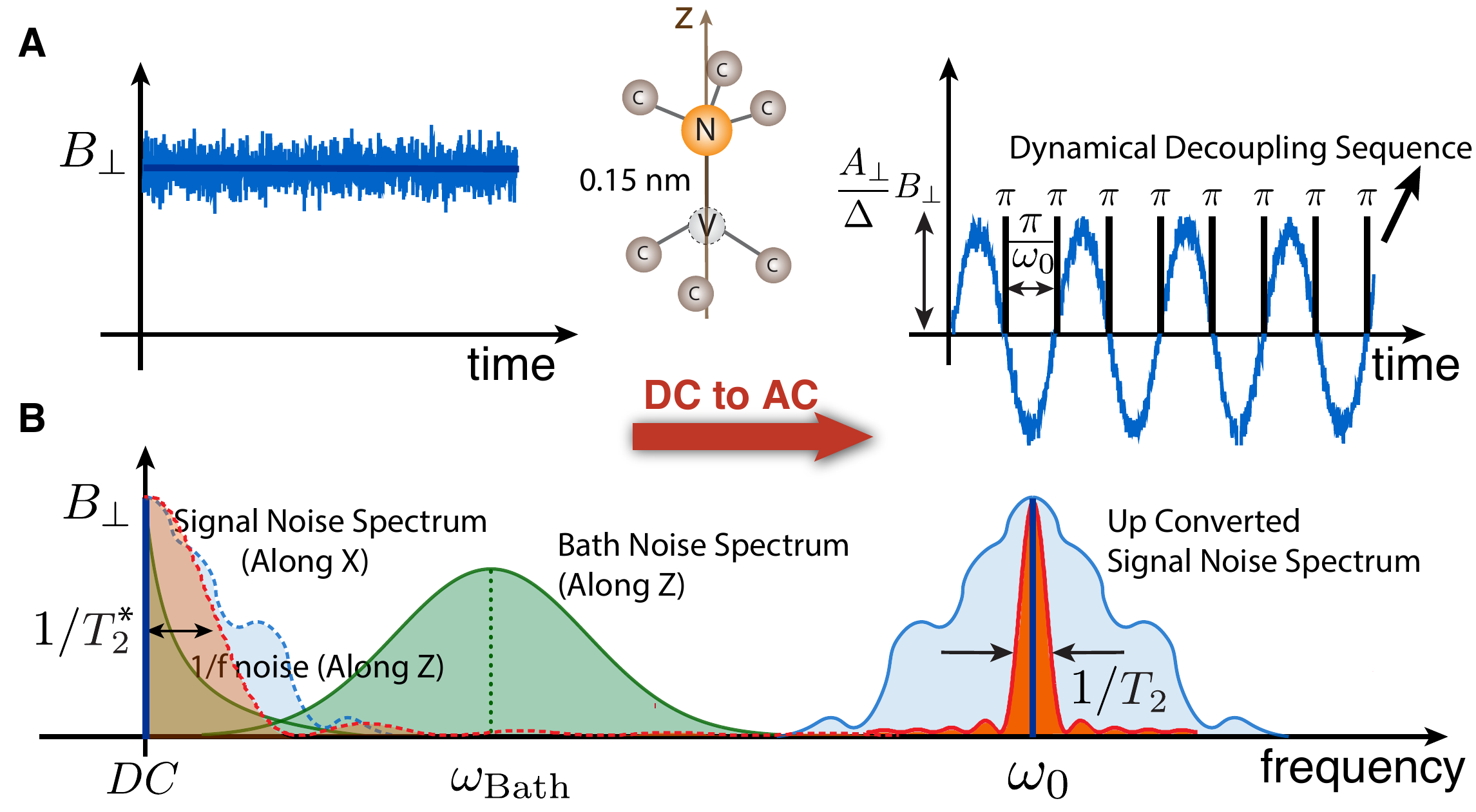}}
  \caption{\textbf{DC magnetometry at T$_2$ via ancilla assisted upconversion.} {(A)} A magnetic field $B_{\pp}$ with low-frequency noise is up-converted by the ancillary $\Ns$ nuclear spin of an NV center  (inset) to an AC signal with frequency $\omega_0$ and amplification factor $A_\perp /\Delta$. Rectification by a DD  sequence (black lines indicate $\pi$ pulses) performs a quantum lock-in detection at $\omega_0$, measuring the DC signal and rejecting noise. {(B)} Frequency-domain picture of the signal frequency upconversion and  noise filtering. The $1/f$ and spin bath noise (green), acting on the NV sensor in the $\bz$ direction,  usually limit interrogation times to $T_2^{\ast}$. We also consider the low-frequency noise carried by the signal (light blue) in the $\bx$ direction before (dashed lines) and after (solid lines) the upconversion. Ancilla assisted upconversion leads to a dual  band-pass filtering of both signal and sensor noise thanks to the narrow DD filter bandwidth,  $\sim1/T_2$, with a consequent gain in magnetometer sensitivity.}
  \label{fig:noise-graph}
\end{figure}

Here we introduce and experimentally demonstrate an ancilla-mediated method to up-convert DC signals to AC signals, prior to  quantum lock-in detection, allowing DC magnetometry  at the $T_2$ limit. While the technique is more general, here we employ a sensor consisting of the electron spin of a single NV center in diamond and an ancillary nuclear spin qubit. 
Coupling to the nuclear spin   up-converts the DC field  to  a frequency that  can be tuned  well beyond the environmental noise spectrum cut-off.
 The AC field is then measured by quantum lock-in-detection (rectification~\cite{Kotler11}). This frequency conversion chain (Fig.~\ref{fig:noise-graph}), from DC$\shortrightarrow$AC$\shortrightarrow$DC, enables  narrowing the noise bandwidth seen by the sensor to within $1/T_2$. 
Our method can thus  be seen as a form of quantum error correction~\cite{Khodjasteh05}, and provides significant gains in sensitivity and noise suppression -- \textit{both} of environment noise and noise carried by the signal to be sensed.

Remarkably, the same frequency upconversion  can also be applied to detect  low frequency (1kHz - 1MHz) AC fields, which are out of reach of conventional dynamical decoupling-based schemes. 
In addition, the long interrogation times make the NV sensor amenable to highly efficient readout techniques,  including charge state readout~\cite{Shields15},  providing significant sensitivity gains. The technique is particularly suited for spin ensemble sensing, by canceling inhomogeneities and providing better sensitivity per unit volume.
Combined with the conventional Ramsey detection of longitudinal fields, our technique can enable sensitive, low noise, full vector nanoscale magnetometry at $T_2$ with a single point defect sensor. 

The paper is organized as follows. Sec.~\ref{sec:protocol} describes the ancilla-based magnetometry protocol followed in Sec.~\ref{sec:sens} by a  sensitivity analysis.  In Sec.~\ref{sec:noise} we detail how our technique, based on frequency upconversion, achieves an effective low-pass filtering of signal noise and band-pass filtering of sensor noise. Sec.~\ref{sec:dynamic-range} evaluates practical issues affecting our technique, in particular the  dynamic range. Finally in Sec.~\ref{sec:applications}, we describe extensions to full vector DC magnetometry and to low-frequency AC magnetometry.

\section{Ancilla-assisted DC field sensing}
\label{sec:protocol}
At the core of our method for frequency up-conversion and lock-in detection is the coherent coupling between a qubit probe and an ancillary qubit system. We assume that the quantum probe $S$ is sensitive to the DC signal we want to detect, but the ancillary qubit $I$ does not couple to it (or the coupling is very weak). The sensor-ancilla Hamiltonian is then:
\begin{align*}
&H=\Delta S_z+\omega I_z	+H_{\textrm{hyp}},\quad\textrm{with} &\\
&H_{\textrm{hyp}}=A_\parallel S_zI_z + A_{\perp}(S_xI_x+S_yI_y)&
\end{align*}
When $\Delta-\omega\gg A_\perp$, the transverse part of the coupling can be neglected ($H_{\textrm{hyp}}\approx A_\parallel S_zI_z $) and the two systems can be easily decoupled, e.g., with a spin echo. However,  an added  transverse magnetic field, $B_\perp$ that couples to the qubit sensor as $\gamma B_\perp S_\perp$,  induces a mixing of the qubit/ancilla energy levels. Under the assumption $\Delta\gg A_\perp$, this second order process produces an effective coupling $H'_{\textrm{hyp}}\approx A_\parallel S_zI_z +\frac{\gamma B_\perp A_{\perp}}\Delta S_zI_x$. Then, in the interaction picture of the ancilla, the last term becomes time-dependent and the qubit sensor effectively sees an AC field $\propto B_\perp\cos(\omega_0 t)$. 
This simple model can be applied to many quantum sensor systems. Here we focus on the electronic spin of a single Nitrogen Vacancy (NV) center in diamond, which has been shown to have AC field sensitivities better than 1nT/$\sq{\textrm{Hz}}$~\cite{Maze08}, and nanoscale spatial resolution~\cite{Mamin13,Staudacher13,Ajoy15}. We consider using a nearby nuclear spin as an ancilla,  for instance the $\Ns$ nuclear intrinsic to every NV center, or a $\Cs$ spin in the first few shells~\cite{Childress06,Shim13}. For this system we have $\Delta=\Delta_0-\gamma_e B_\parallel$ (where $\Delta_0=2.87$ GHz is the NV zero field splitting and $\gamma_e=2.8$ MHz/G is the gyromagnetic ratio of electron), and the effective hyperfine interaction becomes $\frac{\gamma_eB_\perp A_{\perp}}\Delta F S_zI_x$, where $F$ is a factor that depends on the type of ancillary spin  (see Appendix~\ref{app:signal}). Similarly, the frequency $\omega_0$ of the effective AC fields will depend on the spin system considered.

\begin{figure}[t]
  \centering
  {\includegraphics[width=0.5\textwidth]{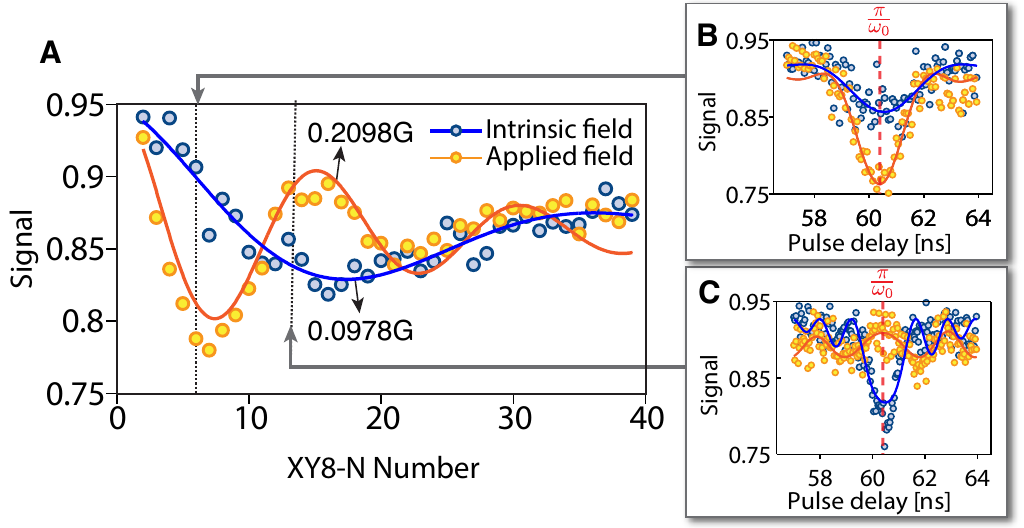}}
  \caption{\textbf{Ancilla assisted DC magnetometry.} {(A)} Magnetometry signal at $\Delta=141$MHz as a function of the pulse sequence length,   under intrinsic misalignment $B_i\approx 0.0978$G (blue points) and  an additional DC  field $B_{\perp}\approx0.2098$G  (orange points). The  field values are obtained from the frequency of the oscillations (solid lines: fit). Each point in (A) is obtained at the optimal sensing point, $\qt=\pi/\omega_0$, from time-domain traces as shown in  {(B)}  and {(C)} for XY8-7 and XY8-12, respectively.  
We employed quantum interpolation~\cite{Ajoy16,SOM} to precisely sample the peaks at 35.7ps (B) and 20.8ps (C). 
}
\label{fig:xy8-time}
\end{figure}

The second-order mixing of the electronic and nuclear levels  can be viewed as an effective frequency up-conversion~\cite{Hill12,Palomaki13, Sridaran12} of the DC field $B_\perp$ to an AC field, $\gamma_e B_\perp (A_\perp F/\Delta)\cos(\omega_0 t)$, at the resonance frequency $\omega_0$ of the  nuclear spin. In other words, the nuclear qubit is analogous to a free running oscillator at a frequency $\omega_0$ set by nature. In the presence of the transverse magnetic field, the NV sensor couples to this oscillator, up-converting the DC field to $\omega_0$. 
This AC field can now be measured using well-known quantum lock-in techniques, based on trains of dynamical decoupling pulses, with pulse spacing set at the effective AC field period.
We can for example use the CPMG/XY8 protocol~\cite{Carr54,Meiboom58,Gullion90} that simultaneously  decouple the NV sensor from external noise, allowing interrogation up to the $T_2$ coherence time~\cite{Kolkowitz12b,Taminiau12}.

The interferometric signal is then given by 
$S=\frac{1}{2}[1+\cos(N\alpha)]$,  where $\alpha$  is the angle between the nuclear Hamiltonians in the two NV manifolds (see \zar{signal}), and $N$ the number of pulses.   
For small $B_\perp $ fields, $\alpha\ll 1$, and to a good approximation the signal is 
\begin{equation}
S= \frac12\left[ 1+\cos\left( \frac{\gamma_eB_\perp A_\perp N}{\omega_0\Delta}F\right) \right].
\label{eq:Signal}
\end{equation} 
This signal is especially strong close to the NV center ground state level anti-crossing (GSLAC,  $B_z\approx 1025$G), where $\Delta\rt 0$. 
The signal slope is  proportional to the hyperfine field $A_\perp $. For $\Ns$, this coupling is $A_\perp = -2.62$MHz and $F=\frac{2\sqrt{2}(Q-A_{\parallel}/2)^2}{(Q-A_{\parallel})Q}\approx 3.1$, as determined from second order perturbation theory.  
Note that while the $\Ns$ spin is a natural choice since it is present in every NV center, and provides the added benefit of high robustness to magnetic fields (see \zar{nitrogen}), an ancillary $\Cs$ spin could provide a wider dynamic range and reduce certain experimental constraints due to its strong hyperfine coupling.
Indeed, for $\Cs$, $F=\frac{4(A_\parallel-\gamma_nB_\parallel)}{2A_\parallel-\gamma_nB_\parallel}\approx 2$, while $A_{\perp}$ could be substantially stronger, $A_\perp > 140$MHz potentially allowing DC field sensing at large values of $\Delta$.

While hyperfine-mediated fields have been observed before~\cite{AbragamBleaney} to boost the Larmor~\cite{Childress06} and Rabi~\cite{Chen15,Sangtawesin15} frequencies of the ancillary nuclear spin, here instead we have employed the hyperfine for a frequency mixing action that aids quantum metrology. 

\begin{figure*}[htb]
  \centering
  {\includegraphics[width=\textwidth]{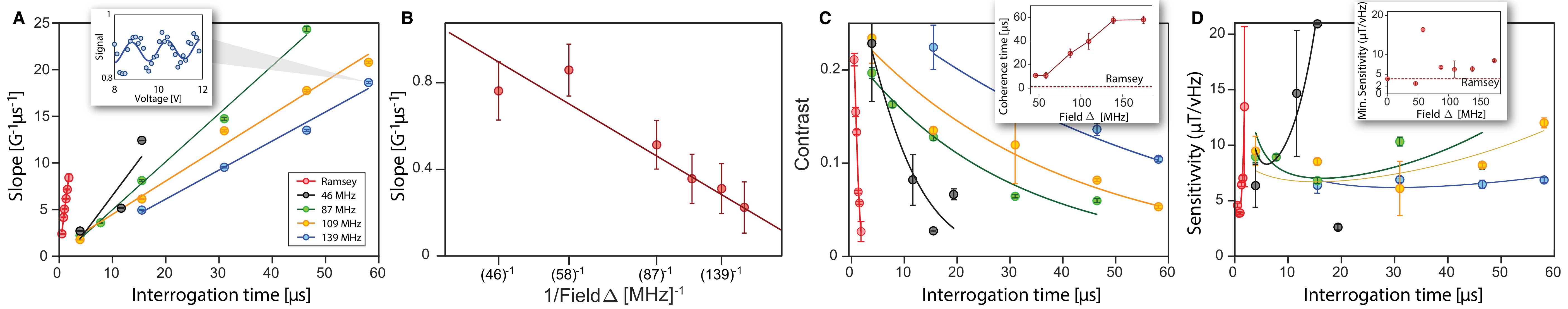}}
  \caption{\textbf{Magnetometer sensitivity.} By sweeping an external DC field $B_e$ we determine the signal sensitivity 
in the linear regime of the magnetometer (Fig.~\ref{fig:sweeping-L}). 
{(A)} Signal slope $\left.dS/dB_e\right|_{B_0}$ as a function of interrogation time for representative values of field $\Delta$. 
Inset: example of magnetometry signal with applied voltage. From the signal Fourier transform we extract  the signal slope (A) and decay(C). 
	The error bars are obtained from a Lorentzian fits of the Fourier transforms.  {(B)} Signal slope per unit time from (A) plotted as a function of $\Delta$, showing the expected $1/\Delta$ dependence. The error bars are residuals to the inverse linear fit. {(C)} Decay of the  signal contrast with increasing interrogation time. In the inset: coherence times extracted from the decay. The Ramsey method (dashed line) is limited by the short $T_2^{\ast}\app 1.15\mu$s, while our ancilla-assisted protocol is limited only by pulse error. {(D)} Sensitivity for the same values of $\Delta$ and interrogation time as in (A-C). Inset:  best experimentally measured sensitivity. Our technique can experimentally achieve sensitivities comparable to the Ramsey method $\app 3.86\mu$T/$\sq{\textrm{Hz}}$, with additional benefits in noise suppression (Fig.~\ref{fig:filter-expt}).
}
\label{fig:sweeping-V}
\end{figure*}

\section{DC Magnetometer Sensitivity}
\label{sec:sens}

\subsection{Theoretical Sensitivity Bounds}
The shot-noise limited sensitivity $\eta$  can be found from the smallest field $\delta B_\perp$ one could measure in the total averaging time, $M(t+t_d)$, needed to perform $M$ measurements with interrogation  time $t$ and dead time $t_d$ (for preparation and readout): $\delta B_\perp\equiv \eta/\sqrt{M(t+t_d)}$~\cite{Taylor08}.
In turns, the minimum field is estimated from its uncertainty at the most sensitive bias point $B_0$~\cite{Itano93},
 $\xd B_\perp  = \left.\frac{\Delta S}{\sqrt MdS/dB_\perp }\right|_{B_0}$, where $\Delta S$ is the spin projection noise $\Delta S = \sq{S(1-S)}$.

From the signal in Eq.~(\ref{eq:Signal}), evaluated at the optimal time $t=N\pi/{\omega_0}$, we obtained the sensitivity
\begin{align}
	\eta=\frac{\omega_0}{A_\perp F} \frac{\Delta\, e^{(t/T_2)^p}}{C\gamma_e N}\sqrt{t+t_d}
	=\frac{\pi\Delta\,e^{(t/T_2)^p}}{C\gamma_eA_{\perp}Ft}\sqrt{t+t_d},
\label{eq:eta}
\end{align}
where we further introduced a signal decay $\propto e^{-(t/T_2)^p}$ and a factor encapsulating readout inefficiencies, $C$. Since the NV spin state is measured  through its fluorescence,  photon shot noise,  finite contrast and photon collection efficiency  degrade the sensitivity by this factor $1/C$~\cite{Taylor08}. 

Since our method exploits DD sequences, the interrogation time $t$ can be extended to $T_2$, and the sensitivity is optimized for $t\approx T_2$.  Assuming $t_d\ll t$~\footnote{$t_d=1.3\mu$s in our experiments, although it could be as low as $\app$350ns}, for $^{14}$N ancilla we obtain
\begin{equation}
	\eta\approx\frac{\pi\Delta}{A_\perp F}\frac{1}{C\gamma_e \sqrt{T_2}}=\frac{\pi\Delta}{A_\perp}\frac{\sqrt{2}Q(Q-A_\parallel)}{4(Q-A_\parallel/2)^2}\frac{1}{C\gamma_e \sqrt{T_2}}.
\label{eq:etafin}
\end{equation}

We can compare this result to the Ramsey sensitivity, $\eta_R=1/(C\gamma_e \sqrt{t})$, where the interrogation time is limited by short dephasing time $t\approx T_2^*$. Even if the ratio $\Delta/A_\perp$ must be kept small (to maintain the perturbative regime as well as achieve low pulse errors, see Sec.~\ref{sec:dynamic-range}), we can achieve an improvement over Ramsey metrology when the ratio $T_2/T_2^*$  is large, as it is often the case (e.g., $T_2/T_2^*\sim10^3$ in NV systems).

\subsection{Experimental Sensitivity}
To experimentally determine the sensitivity for our technique, we measure the signal slope $\left.dS/dB_\perp \right|_{B_0}$ for a $\Ns$ ancilla by sweeping 
an external field $B_e$. The field is produced by 
an external voltage $V$ (Fig.~\ref{fig:sweeping-V}), and it is simply proportional to V, with $B_e/V =$ 0.0407$\pm$ 0.0006 G/V  (obtained from the experiments described in Fig.~\ref{fig:sweeping-L}). 
We operate in the optimal dynamic range regime of the magnetometer (see \zar{intrinsic}), where the field to be measured, perpendicular to the N-V axis, is  proportional to the  external field $B_\perp \propto B_e$.

As shown in Fig.~\ref{fig:sweeping-V}.(A-C), at the optimal bias point the signal slope is $\left.dS/dB_\perp \right|_{B_0}=  e^{- \lb t/T_2\rb ^p}\frac{ \gamma_e A_\perp F \,t}{2\pi\Delta}$, i.e. it varies linearly with the interrogation time, is inversely proportional to the deviation $\Delta$ from the GSLAC, and is reduced by the signal decay. For a more accurate estimation of the slope, we measured the signal over a larger range of applied fields and used its Fourier transform to independently determine the signal decay (from the contrast, Fig.\ref{fig:sweeping-V}.C) and its dependence on $B_\perp$ (from the frequency (slope), Fig.\ref{fig:sweeping-V}.A).
 
For comparison, we also plot the performance of conventional Ramsey magnetometry experiments under the same conditions (Fig.~\ref{fig:sweeping-V}) for the measurement of the \textit{longitudinal} magnetic field, $\cot(\theta)B_\perp$, where $\theta=39.32^{\circ}$ is evaluated from the experiments shown in Fig.~\ref{fig:sweeping-L}.
In Fig.~\ref{fig:sweeping-V}{(B)}, one can clearly discern the $1/\Delta$ dependence of the signal slope showing  very good agreement with theory. From Fig.~\ref{fig:sweeping-V}{(C)}, it is evident that in our technique the NV sensor is better protected against noise and much longer interrogation times are possible than in the Ramsey experiment,  for  in the presence of the external field,  $T_2^*\app 1.16\mu$s (inset of Fig.~\ref{fig:sweeping-V}{(C)}). 

We measured the signal noise $\Delta S/C\app 17.27$ through a binning of photocounts, and estimated the magnetometer sensitivity for different values of $\Delta$ and interrogation times (Fig.~\ref{fig:sweeping-V}{(D)}). 
At $\Delta=139$MHz the sensitivity of our ancilla assisted protocol $\app 6.02\mu$T$/\sq{\textrm{Hz}}$ approaches that of the Ramsey technique $\app 3.86\mu$T$/\sq{\textrm{Hz}}$ (inset of Fig.~\ref{fig:sweeping-V}{(D)}). 
As explained in \zar{intrinsic}, pulse errors limited the achievable interrogation times to $t\approx 60\mu$s.
The errors are higher than usual partly because of the high duty cycle (the ratio between the pulse duration and the period)  of the DD sequence ($t_\pi/\tau\app 0.41$ at $\tau=\pi/\omega_0\app 121$ns and $\pi$-pulse length $t_\pi=50$ns), 
and due to the residual intrinsic misalignment of about $ 0.3$G  (\zar{intrinsic}). Indeed, we achieve a better  coherence time, $T_2>350\mu$s,  at lower pulse duty cycles $\app 0.047$.
 Hence, by employing faster pulses, working at high nuclear harmonics (Fig.~\ref{fig:harmonic-plot}), and using magnetic shielding to remove intrinsinc misalignment~\cite{Kornack07}, we should be able to   improve the pulse quality, achieve longer  interrogation times, and  outperform the Ramsey sensitivity. An alternative strategy (\zar{HH}) is to use a spin-lock pulse sequence that can achieve long coherence times ($T_{1\rho} \gg T_2^*$).

Our protocol could provide further sensitivity gains when combined with the recently developed readout via NV spin-to-charge mapping followed by charge state readout (CSR)~\cite{Shields15}. 
CSR can dramatically improve the C-factor in magnetometry at the cost of an increased dead time $t_d$. For the Ramsey technique, $t_d$ is a substantial fraction of the interrogation time and CSR leads to little overall gain in sensitivity. In contrast,  long interrogation times, as in our method, allow to reap the benefits of CSR, as even longer dead-times are still just a fraction of the interrogation times. For example,  for $t_{d}\app T_2/4$, $C_{\textrm{CSR}}\shortrightarrow 10C$ for $T_2\app 1$ms~\cite{Jaskula15}. For the interrogation times in our system, $T_2\app 60\mu$s, we have $C_{\textrm{CSR}}\app 2C$~\cite{Jaskula15} and the sensitivity can be improved to $\app 3.01\mu$T$/\sq{\textrm{Hz}}$. 

We also note that for DC magnetometry with an ensemble of NV sensors, performing the experiments  at the GSLAC produces an additional boost in $C$-factor by $\app$1.43, since the photoluminescence background from off-axes families of NV centers,  which do not contribute to the signal, is effectively quenched by a factor $\sim $0.6. (See \zar{magnet}, Fig.~\ref{fig:magnet-alignment} showing this effect.) 

Finally, our technique  leads to additional gains in sensitivity due to superior suppression of signal and sensor noise as we  now explain.

\section{Dual suppression of Signal and Sensor Noise}
\label{sec:noise}
Since our magnetometry method exploits an ancilla-assisted frequency up-conversion of the DC field, it enables an enhanced rejection of noise and consequently interrogation  times approaching $T_2$. 
The noise rejection applies to two main sources of noise (Fig.~\ref{fig:noise-graph}) -- the noise due to the local environment, always affecting the sensor (\textit{sensor noise}), and the additional noise introduced by  the magnetic field being sensed (\textit{signal noise}).  
The signal noise is in the transverse ($\bx$) direction and centered around zero or low frequency, while the sensor noise is primarily in the $\bz$ direction and centered about $\omega_{\textrm{Bath}}$. As we describe in detail in \zar{noise-filter}, our technique enables \textit{low-pass} filtering of signal noise, and \textit{band-pass} filtering of sensor noise. 
The filtering action is due to the combination of two factors, DD-sequence filtering with  ancilla-assisted frequency upconversion. DD sequences can indeed be interpreted as creating narrow-band filters~\cite{Biercuk11,Ajoy11, Alvarez11}, centered around the inverse pulse period and with rejection and band-pass width inversely proportional to the number of pulses. In Ramsey magnetometry, this filter is centered around zero frequency, where the dominant $1/f$ sensor noise  is maximum, and it has the widest band-pass width. 
In analogy with conventional lock-in-amplifiers~\cite{hill15}, in our method the ancilla performs a frequency upconversion of the DC field to $\omega_0$, allowing us to lock into this frequency with a DD sequence and to effectively \textit{narrow} the noise bandwidth seen by the sensor. 
In addition, since the suppression of sensor noise allows longer interrogation times, the band-pass of the filter can be made very narrow by increasing the pulse number, thus becoming very effective in suppressing the signal-noise (which undergoes a frequency upconversion in the same way as the DC signal field).
\begin{figure}[t]
  \centering
  \includegraphics[width=0.5\textwidth]{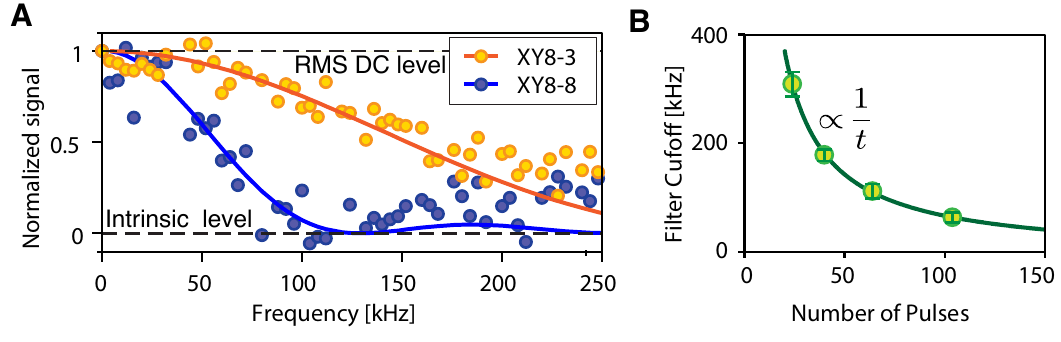}
  \caption{\textbf{Experimental low-pass filtering of signal noise.} {(A)} Signal depth of two XY8-N (yellow, $N=3$, blue $N=8$) experiments  at $\Delta=$141 MHz as a function of the frequency of an applied  magnetic field. The signal  directly maps the filter function of the protocol. Three sets of data are taken concurrently: signal  in the presence of an applied AC magnetic field, signal  without AC magnetic field (which sets the intrinsic level), and signal  in the presence of a  DC magnetic field . The first set of  data is normalized against these two levels. Solid lines are the theoretical low-pass filters (\zr{filterL}),  showing very good agreement with the experimental data. {(B)} Cut-off frequency of the filter, defined as the first zero crossing of the experimental signal in (A). The filter becomes narrower with increasing interrogation times (pulse number) as expected from theory (solid line), with the lowest measured point corresponding to a cutoff of 63.4kHz. In contrast, the filter for the Ramsey technique is much broader, with cutoff $1/T_2^*\app$ 862.0kHz. The error bars are obtained from fits of the theoretical low-pass filters.}
\label{fig:filter-expt}
\end{figure}

Qualitatively similar gains in interrogation times could be achieved by exploiting an ancilla to perform quantum error correction~\cite{Kessler14,Arrad14,Dur14}. Whereas these techniques could  potentially go even beyond $T_2$, provided feedback correction is applied faster than the error rate affecting the sensor, in our protocol the ancilla need not be initialized nor actively manipulated,  greatly reducing the experimental resources required.

In Fig.~\ref{fig:filter-expt} we experimentally demonstrate the signal noise filtering ability of our protocol, by {mapping} its noise filter lineshape. We reproduce the expected low-pass frequency behavior (\zar{noise-filter}), where the cutoff decreases inversely with the total interrogation time $t$. We demonstrate a low-pass signal noise cutoff of 63.4kHz for XY8-13, in contrast to the much larger Ramsey cutoff of 869.6kHz. 
In principle it could be possible to reduce the signal noise with additional hardware~\cite{Vion95}, however this filter stage would inevitably increase the sensor standoff from the signal source, and thus reduce the sensitivity. Our method is then equivalent to introducing an intrinsic tunable low-pass signal filter, with just a 0.15nm additional footprint.

Signal noise rejection leads to a practical improvement in the minimum field that can be sensed. Indeed, $\delta B_\perp$ decreases with the number $M$ of measurements as $1/\sqrt M$, as can evaluated by the sensor Allan deviation. 
However, even assuming not to be limited by drifts in the experimental setup, this shot-noise regime is bounded by the signal noise spectrum  cut-off $1/\omega_n$. Hence, in the regime $T_2^\ast<1/\omega_n<T_2$, a suppression in signal noise allows a gain in achievable shot-noise repetition number $n_r$, improving the sensitivity by a factor $\sqrt{T_2\omega_n}\leq\sqrt{{T_2}/{T_2^\ast}}$ over the conventional Ramsey technique. 

Suppression of sensor noise is particularly beneficial 
 for ensemble magnetometer operation, where inhomogeneities more severely affect $T_2^*$-limited protocols.
For $N_v$  sensors per unit volume, the suppression of spin bath sensor noise via DD~\cite{Wyk97,Bauch15} enables keeping $N_v/T_2\app 1$, and hence an approximately constant sensitivity per unit volume.

\section{Dynamic Range and Practical limitations}
\label{sec:dynamic-range}
Various experimental issues affect the ultimate performance of our method, determining the practically achievable sensitivity and the optimal operating regime. Here we consider  limitations of our current experiments, as well as more general bounds and strategies to overcome them.
\begin{figure}[t]
  \centering
  {\includegraphics[width=0.5\textwidth]{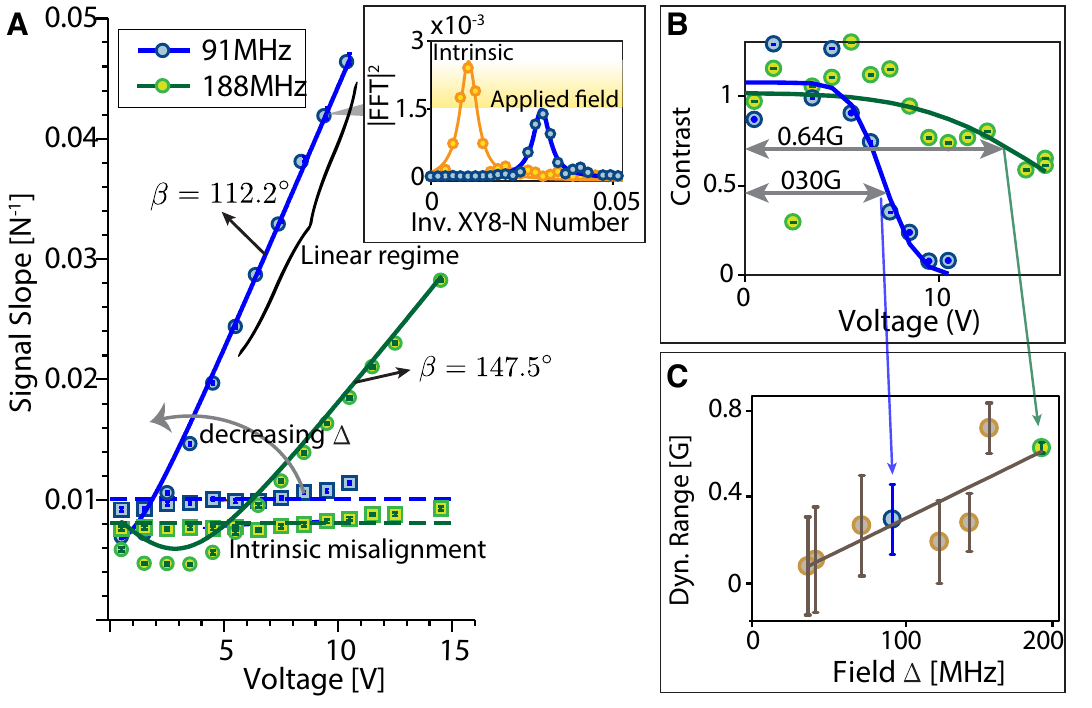}}
  \caption{\textbf{Optimal  operating range.} 
{(A)} From the Fourier transform of the signal as a function of XY8-N cycle number N, we extract  the signal frequency $dS/dN$ as a function of the applied transverse field $B_e$, for two values of the longitudinal magnetic field, $\Delta$. Inset: representative signal Fourier transform for two magnetic fields (intrinsic misalignment $B_i$, and total transverse field $B_{\perp}$).
The solid lines are fits following $B_{\perp}=\sqrt{B_i^2+B_{e}^2+2B_iB_{e}\cos\beta}$ (Fig.~\ref{fig:parabola}). We highlight the optimal (linear) operating regime where $B_e\gg B_i$. We also extract the values of $\beta$, in-plane angle between $B_e$ and $B_i$. 
The squares are the measured values of $B_i$, and dashed lines represent the mean $B_i$ (blue and green correspond to intrinsic misalignments of 0.084G and 0.142G). Error bars are obtained from a Lorentzian fit of the Fourier transform peaks. For field values closer to the GSLAC (decreasing $\Delta$) the signal slope $d^2S/dB_edN$ increases, leading to increased magnetometer sensitivity (see Fig.~\ref{fig:sweeping-V}). The decrease in signal contrast in the inset is an indication of finite sensor dynamic range. The contrast is plotted in {(B)} as a function of the applied field. We fit the data to sigmoid functions (solid lines) from which we extract the sensor dynamic range.  Note that the normalization is performed with respect to the concurrently measured signal under intrinsic misalignment.  
{(C)} Dynamic range for different values of $\Delta$, indicating that it decreases with smaller $\Delta$. Solid line is a linear fit.} 
\label{fig:sweeping-L}
\end{figure}

The dynamic range is a function of the deviation from the GLSAC,  decreasing with smaller $\Delta$. In our experiments, a chief factor affecting the dynamic range  is the residual field from  Earth, $B_i$, which sets the lower limit of the dynamic range (see \zar{intrinsic}). Since this problem could be resolved simply by magnetic shielding, routinely used in precision magnetometry, here we focus instead on the upper limit to the operating range, i.e. the maximum field $B_\perp $ that can be applied to our sensor without loss of sensitivity. 

First,  the presence of $B_\perp $ causes a small tilt of the initial state prepared after the first $\pi/2$ pulse (assumed to be ideal until now) leading to a loss in signal contrast -- and hence magnetometer sensitivity -- by an amount $\Delta^2/(\Delta^2 + B^2_\perp)$. This effect is negligible for $B_\perp \ll \Delta$ and it can be mitigated even if the field to be measured is large, provided it is known to some precision.
For a known bias point of magnetometer operation $B_{0_\perp }$, one can \textit{compensate} the tilt with a suitable preparation pulse with flip angle $\frac{\pi}{2}-\arctan(B_{0_\perp} /\Delta)$. 

A second factor, affecting both dynamic range and sensitivity,  is the lower quality of the pulses close to the GSLAC, which limits the efficiency of DD sequences and thus the coherence time. To obtain good pulses, we need a sufficiently large Rabi frequency $\Omega$, such  that $B_\perp \ll\Omega$. However, the Rabi frequency has also other constraints.  
When $\Delta\to 0$, the pulses degrade due to the breakdown of the rotating wave approximation (RWA). It is not possible to simply reduce the Rabi frequency, as the minimum allowed Rabi frequency $\Omega$ is constrained by the pulse separation time, in turn set by $\omega_0$, $\Omega = \omega_0$. 
Then, imposing $\Omega< \Delta/5$ for the RWA to hold, sets a bound on the closest approach to the GSLAC. For example, $\Delta\gtrsim 5\omega_0\app 20$ MHz for the $\Ns$ nuclear ancilla used in our experiments. 
 In addition,  even within the RWA  the pulse quality is degraded due to off-resonance effects caused by $B_\perp $ -- with a pulse error (infidelity with respect to an ideal $\pi$-pulse~\cite{NielsenFid}) $\sim B_\perp^2/2\Omega^2$. 

These errors are compounded upon increasing the number of pulses, especially at high duty cycles, leading to a loss of magnetometer sensitivity. However, the knowledge of the operating bias point $B_{0_\perp }$ would allow pulses to be suitably calibrated and optimized, even going beyond the RWA~\cite{Hirose15x,Fuchs09,Scheuer14}. Furthermore, the use of higher order pulse sequences, widely developed in the NMR community~\cite{Alvarez10b,Souza11}, can suppress these errors to a large extent. 

The presence of a finite dynamic range is evident in the decay of the curves in Fig.~\ref{fig:xy8-time}, the pulse error compensation of XY8 sequences being slightly imperfect at high pulse number. We perform a detailed experimental characterization (Fig.~\ref{fig:sweeping-L}
) of the sensor dynamic range by determining the loss in signal contrast  upon sweeping the external magnetic field (voltage), for increasing XY8 cycle number, and for different values of $\Delta$.  Fig.~\ref{fig:sweeping-L}{(B)} shows a representative example of this behavior -- at high applied voltages, there is a sharp loss of contrast beyond the sensor dynamic range. By performing a fit to a sigmoidal function $\sim \frac{\exp(-B_e - B_{t})}{[1+ \exp(-B_e - B_{t})]}$, we extract the turning point $B_{t}$ that serves as a good measure of the sensor dynamic range. From Fig.~\ref{fig:sweeping-L}{(C)} it is evident that the dynamic range  decreases as one approaches the GSLAC.

\section{Applications \& Outlook}
\label{sec:applications}
\subsection{Nanoscale vector magnetometry}
Combining our technique with conventional Ramsey DC magnetometry can be the basis 
for a protocol for \textit{vector} magnetometry of DC fields~\cite{Pham11,Geiselmann13} with a \textit{single} NV sensor,  with  nanometer spatial resolution. Acquiring vectorial information about magnetic and electric fields would be of great benefit in many practical applications, from bioimaging (e.g. for  neuronal magnetic fields~\cite{Barry16}) to materials science.
In the vector magnetometry protocol,  Ramsey magnetometry  measures the longitudinal ($B_{\parallel}$) component of the field, while  our ancilla-assisted technique detects its transverse  component ($B_{\perp}$). For instance from the experiments shown in Fig.~\ref{fig:sweeping-V}, we  determine the polar angle of the field to be $\xt=39.32^{\circ}$ ($\tan(\xt) = B_{\perp}/B_{\parallel}$). In order to further determine the direction of $B_{\perp}$ in the transverse plane, one can apply a  bias transverse field in a known direction (say $\bx$). By measuring again the total transverse field with the ancilla-assisted technique, one can  extract  the direction  of the unknown $B_\perp$ (its azimuthal angle $\phi$ with respect to  $\bx$). Note that this is equivalent to experiments described in Fig.~\ref{fig:sweeping-L}.

\begin{figure}[t]
  \centering
  {\includegraphics[width=0.5\textwidth]{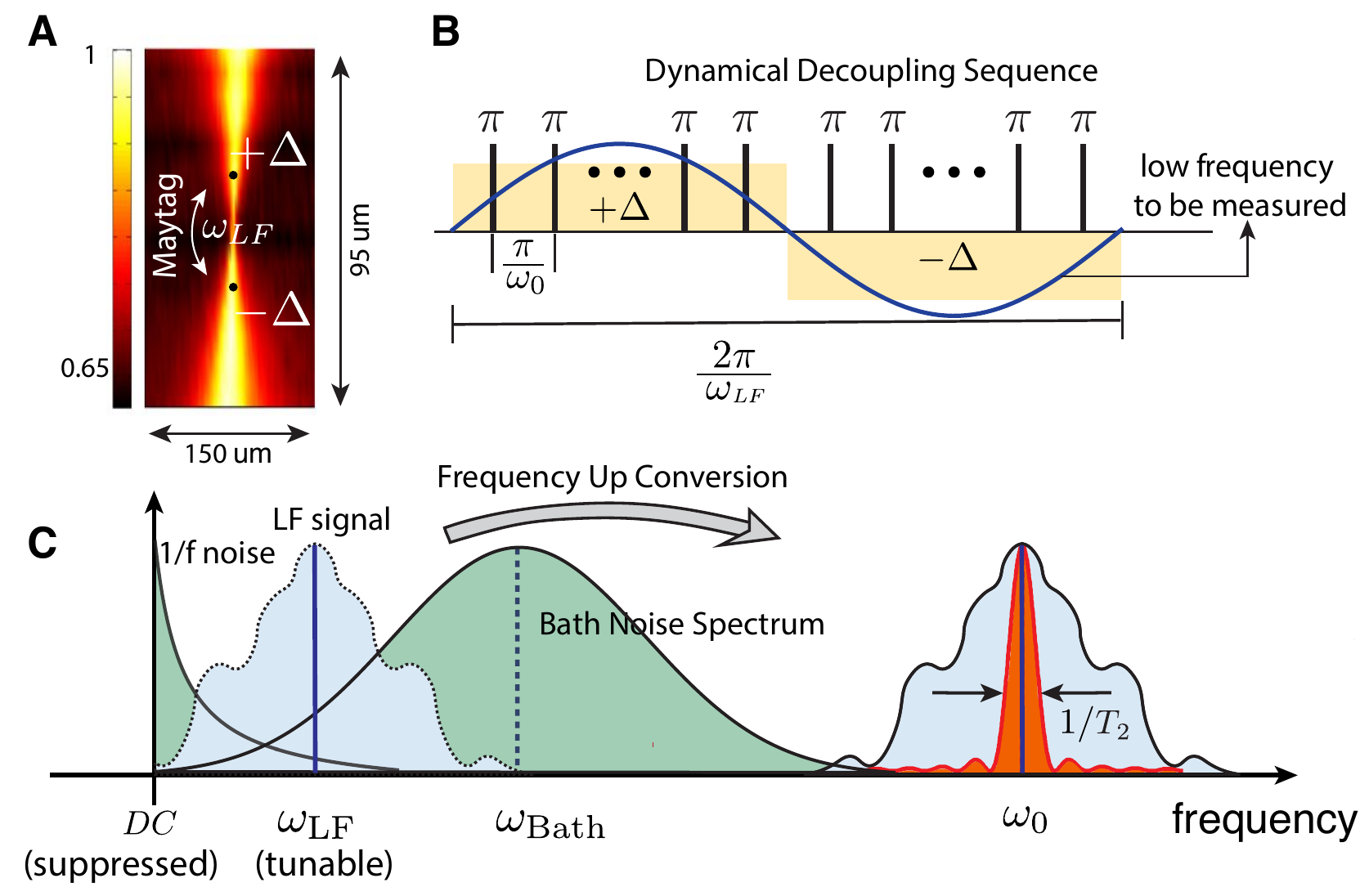}}
  \caption{\textbf{Maytagging for low frequency AC magnetometry.}  Schematic illustration of the proposed method for sensing low-frequency AC fields. {(A)} NV photoluminescence in an XZ magnet scan close to the GSLAC (see Fig.~\ref{fig:magnet-alignment}). (B) Pulse sequence for low-frequency magnetometry, where we alternate (maytag) the field between two values of $\Delta$ symmetric with respect to the GLSAC, at a rate $\omega_{\textrm{LF}}$ (as low as $1/T_2$).  {(C)} Frequency domain representation, wherein the signal at $\omega_{\textrm{LF}}$ is up-converted to $\omega_0$ and detected via dynamical decoupling.}
	\label{fig:maytag}
\end{figure}

\subsection{Low Frequency AC Magnetometry}
\label{app:maytagging}
Variants of our technique can also be applied to  sense  low frequency fields that are outside of range of usual DD sensing. This would be an invaluable tool for measuring  several ubiquitous fields of experimental significance that have  low frequency (30kHz - 1MHz). For instance, this a common occurrence for impulse response fields from magnetic materials~\cite{Baker12}, where the repetition rate is often material-limited to $\omega_{\textrm{LF}}<1$MHz. 
Unfortunately the conventional echo-based AC field sensing techniques require the pulsing frequency to be exactly matched with the frequency one wants to sense, which is not possible when $\omega_{\textrm{LF}}\ll 1/T^{\ast}_2$ due to sensor dephasing. Hence the low frequency regime is a ``blind-spot'' for conventional magnetometry~\cite{Yan15}. 

\begin{figure}[t]
  \centering
  {\includegraphics[width=0.48\textwidth]{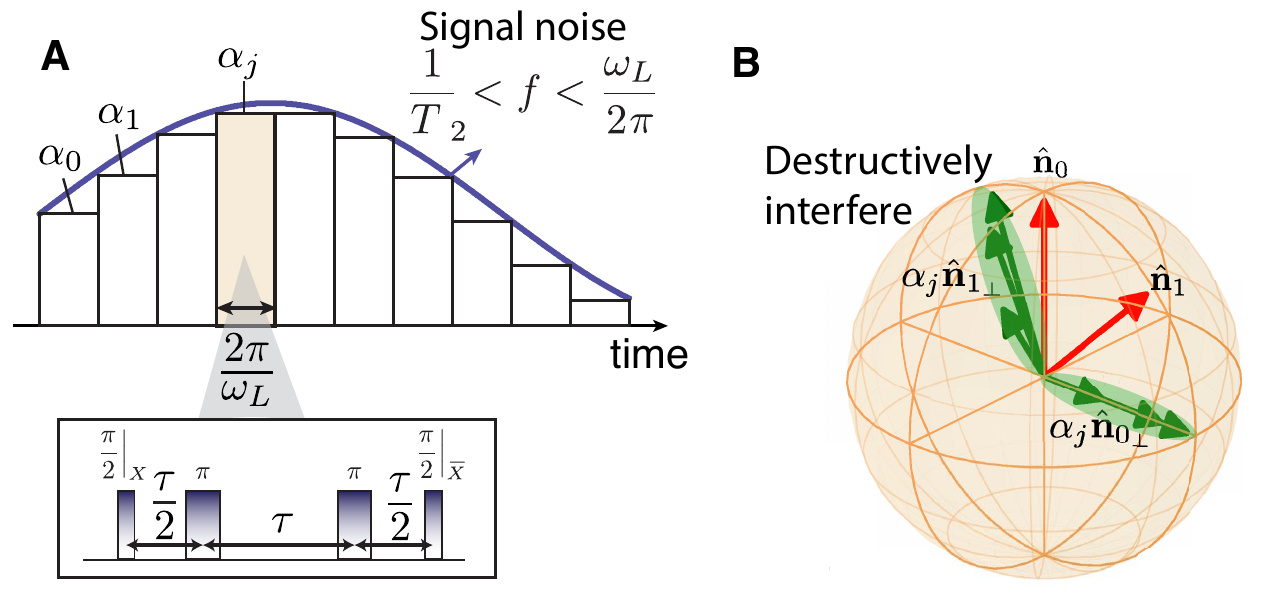}}
  \caption{\textbf{Low frequency noise rejection in ancilla-assisted magnetometry.} Schematic derivation of the low-pass signal noise filter inherent to our protocol. {(A)} Consider a low frequency magnetic noise tone of frequency $f\ll \omega_0$. Each rectangular block corresponds to a CPMG pulse sequence (inset) of total time $2\qt=2\pi/\omega_0$. To a good approximation the signal results from the combined effect of the tilt angle $\alpha_j=B^{(j)}_\perp A_\perp F/\Delta$, discretized over the $L$ CPMG blocks. {(B)} Geometric visualization of filtering action. Here $\bn_{0}$ and $\bn_{1}$ (red arrows) are the axes of nuclear rotation. For a \textit{static} DC field, the effective vectors are perpendicular to the original axes and anti-parallel leading to a strong signal through  constructive interference. However in the presence of low-frequency signal noise, these open up a fan of vectors (green arrows), for instance $\alpha_j\bn_{1^{(j)}_\perp }$, over which the effective propagator needs to be evaluated. This leads to destructive interference, and consequently signal noise suppression that increases with number of pulses. }
  \label{fig:sine-wave}
\end{figure}

Due to the narrowband low pass filtering behavior of the sensor, we cannot directly use our method to measure low frequency AC field. However, an extension of the method, inspired by field reversals employed in MEMS technology~\cite{prikhodko13}, can detect slowly varying fields.
 We alternate (\textit{maytag}) the applied longitudinal magnetic field between two values  that reverse the sign of  $\Delta$ (Fig.~\ref{fig:maytag}), with the maytagging frequency equal to the frequency of interest, $\omega_{\textrm{LF}}$. In the experiments, this can be  achieved for instance by the application of a fast switching control voltage inducing a $B_z$ field, in addition to a fixed bias field  at the GSLAC condition, $\Delta=0$. 
Maytagging leads to  \I{amplitude} modulation of the field to be sensed, $B_\perp(t) A_\perp /[\Delta\cos(\omega_{\textrm{LF}} t)]$, and only frequencies in $B_\perp(t)$ that match the maytagging frequency contribute to the signal. All other components, including the DC component, are filtered out in the same manner as the filtering of signal noise in Fig.~\ref{fig:filter-expt} (see \zar{noise-filter} for a detailed derivation).

Crucially, this sensing frequency is now under experimental control, and can be as low as $1/T_2$ with no loss in sensitivity. 
Importantly, this enables moving away from the zero frequency point where the $1/f$ noise has a  significant contribution. In summary therefore, a tunable frequency up-conversion from a low $\omega_{\textrm{LF}}$  to a high-frequency $\omega_0$ using our method, allows one to decouple the pulsing frequency from the sensing frequency in conventional AC magnetometry.  

\section{Conclusions} 
 We have demonstrated a new modality for high sensitivity magnetometry through a  frequency upconversion mediated by an ancillary nuclear spin, combined with quantum lock-in detection,  allowing low noise DC magnetometry at the $T_2$ limit. In contrast, conventional DC magnetometry is limited by the much shorter dephasing time $T_2^*$.
The technique leads to an enhanced suppression of noise, both of the external noise on the sensor as well as the noise associated with the signal being sensed. For sensors based on NV centers in diamond, this could enable substantial gains in DC sensitivity over the conventional Ramsey method, especially when coupled with spin-to-charge conversion and readout. In conjunction with Ramsey spectroscopy, our technique enables full vector DC magnetometry at the nanoscale. Similar upconversion ideas can be applied to the sense low frequency AC fields. We anticipate applications of our sensing technique in many areas of biology and condensed matter physics, and for the transduced sensing of rotations, pressure and electric fields.\\

\textit{Acknowledgments --} We gratefully acknowledge discussions with S. Bhave, R. Walsworth, F. Jelezko, J. Barry, D. Glenn, L. Marseglia, U. Bissbort, J.C. Jaskula and K. Saha. This work was supported in part by the NSF CUA and the U.S. Army Research Office.

\begin{appendices}



\begin{figure}[t]
  \centering
  {\includegraphics[width=0.45\textwidth]{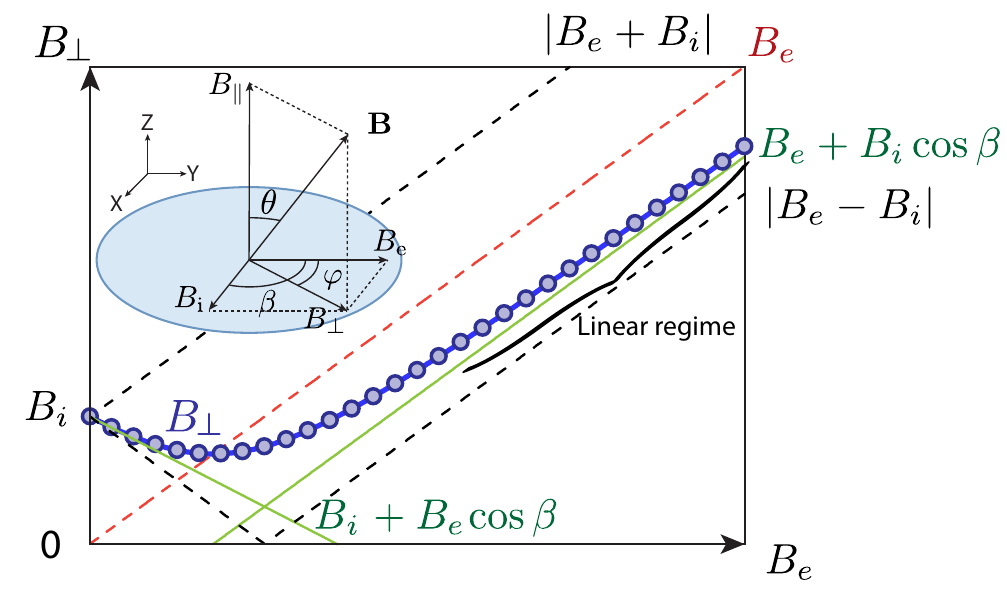}}
  \caption{\textbf{Geometric interpretation of optimal magnetometer operating regime.} The effective field $|B_{\perp}|$ is the vector sum (inset) of the external field to be measured $B_e$ and the intrinsic misalignment $B_i$, separated by the  angle $\beta$ (here $\beta=3\pi/4$).  Green lines denote asymptotes in the extremal regimes $B_e\gg B_i$ and $B_e\ll B_i$, while black dashed lines are the extreme values of $|B_{\perp}|$, for $\beta=0,\pi$. In the regime $B_e\gg B_i$, the total field grows linearly with $B_e$, and there is no loss of sensitivity due to intrinsic misalignment. This is the optimal operating regime for the NV sensor (see also Fig.~\ref{fig:increasing-misalignment}). 
The experiments of Fig.~\ref{fig:sweeping-L}, where the signal slope measures $B_{\perp}$, clearly reproduce the  behavior depicted here.  }
\label{fig:parabola}
\end{figure}

\section{Second order process producing magnetometry signal}
\label{app:signal}

We consider more in detail the second order process that leads to the DC magnetometry signal close to the GSLAC. The effective Hamiltonian of the coupled NV-$^{14}$N ancilla system is, $H= H_0 + V$
where, 
\begin{equation}
H_0=\Delta_0 S_z^2 + B_z (\gamma_e S_z + \gamma_n I_z) + Q_0 I_z^2 + A_\parallel S_z I_z,\end{equation} 
is the dominant interaction, and 
\begin{equation}
V=\gamma_e B_{\pp}S_x + \frac{A_{\perp}}{2}(S_+ I_- + S_- I_+),
\end{equation} 
is the perturbation. 
We define $\Delta\equiv\Delta_0-\gamma_e B_z $ and $Q\equiv Q_0+\gamma_nB_z$ (note that $\gamma_e$ is positive, $Q_0$ and $\gamma_n$ are negative). Both NV electronic spin and $^{14}$N nuclear spin are spin-1 systems, but for simplicity we only consider the $m_s=\ket{0},\ket{-1}$ manifold and $m_N =\ket{+1},\ket{0}$ manifold, where the evolution occurs. Through a Schrieffer-Wolff transformation \cite{Schrieffer66,BravyiSW} the perturbed Hamiltonian $H_0+V$ is transformed to a block-diagonal form \begin{equation} H_{\R{eff}} = \ket{0} \bra{0}H_{\ket{0}}+\ket{-1}\bra{-1}H_{\ket{-1}}, \end{equation} where $H_{\ket{0}}=\mathbf{n}_0\cdot\hat{\sigma}$, $H_{\ket{-1}}=\mathbf{n}_1\cdot\hat{\sigma}$, with 
\begin{align}
	\mathbf{n}_0 = &\left[ \frac{1}{2\sqrt{2}}A_{\perp}\gamma_eB_{\perp}\left(\frac{1}{A_{\parallel}-\Delta}+\frac{1}{A_{\parallel}-Q-\Delta}\right),0,\right.\nonumber \\
	&\left. \frac{1}{2}\left(Q+\frac{A_{\perp}^2}{Q-A_{\parallel}+\Delta}\right) \right],\\
\mathbf{n}_1=& \left[\frac{1}{2\sqrt{2}}A_{\perp}\gamma_eB_{\perp}\left(\frac{1}{\Delta}+\frac{1}{Q-A_{\parallel}+\Delta}\right),0,\right. \nonumber\\
&\left. \frac{1}{2}\left(Q-A_{\parallel}+\frac{A_{\perp}^2}{Q-A_{\parallel}+\Delta}\right)\right]. 
\end{align} 
Since the norms of $\mathbf{n}_0$ and $\mathbf{n}_1$ are close,  to a good approximation the ancilla spin resonance frequency is 
$\omega_0 =\frac{1}{2}(Q+\frac{A_{\perp}^2}{Q-A_{\parallel}+\Delta}) +\frac{1}{2}(Q-A_{\parallel}+\frac{A_{\perp}^2}{Q-A_{\parallel}+\Delta}) = Q-\frac{A_{\parallel} }2+\frac{A_{\perp}^2}{Q-A_{\parallel}+\Delta}$. 
The angle between $\mathbf{n}_0$ and $\mathbf{n}_1$ is $\alpha \approx \sin \alpha = \frac{\mathbf{n}_0\times\mathbf{n}_1}{|\mathbf{n}_0||\mathbf{n}_1|}\approx\frac{\gamma_eB_{\perp}A_{\perp}}{\Delta\omega_0}F$, where we defined $F\equiv \frac{2\sqrt{2}(Q-A_{\parallel}/2)^2}{(Q-A_{\parallel})Q}$.

 In a CPMG sequence, the delay between pulses $\qt$ is typically swept (Fig.~\ref{fig:xy8-time}), and when the delay matches the condition $\qt=\pi/\omega_0$, the signal is  interferometrically obtained as an effective overlap of two propagators conditioned on the NV state, $\mU_{\ket{0}}= \exp(-i\alpha\bs\cdot\bn_{1_\perp })$ and $\mU_{\ket{1}}= \exp(i\alpha\bs\cdot\bn_{0_\perp })$, and is of the form $S=  \frac{1}{2}\lsb 1+ \frac{1}{2}\textrm{Tr}\lcb\textrm{Re} \lb \mU_{\ket{0}}^{\dg}\mU_{\ket{-1}}\rb \rcb\rsb$. 
With increasing interrogation times, effectively achieved through the application of larger number $N$ of  $\pi$-pulses, there is a linear phase accumulation in the interferometric detection~\cite{Ajoy16}, $\Tr{\mU_{\ket{0}}^{N/2}\mU_{\ket{-1}}^{{N/2}\dg}} =2\cos^2(N\alpha/2) -
  2(\bn_{0_{\perp }}\cdot \bn_{{1}_{\perp }})\sin^2(N\alpha/2)$.  
The last term is responsible for the growing signal strength with pulse number, giving $S=1-\sin^2 (N\alpha/2) \cos^2(\alpha/2)\app  \frac{1}{2}(1+\cos(N\alpha))=\frac{1}{2}(1+\cos(N \frac{\gamma_eB_\perp A_\perp }{\Delta\omega_0}\frac{2\sqrt{2}(Q-A_{\parallel}/2)^2}{(Q-A_{\parallel})Q}))$ (see Fig.~\ref{fig:xy8-time}).
 
A similar derivation can be made for the NV-$^{13}$C coupled system. We obtain $F=\frac{4(A_\parallel-\gamma_nB_\parallel)}{2A_\parallel-\gamma_nB_\parallel}\approx 2$ and we consider $\omega_0=\gamma_{n}B_\parallel$ for maximum signal contrast in the sensitivity calculation.

\begin{figure}[t]
  \centering
  {\includegraphics[width=0.45\textwidth]{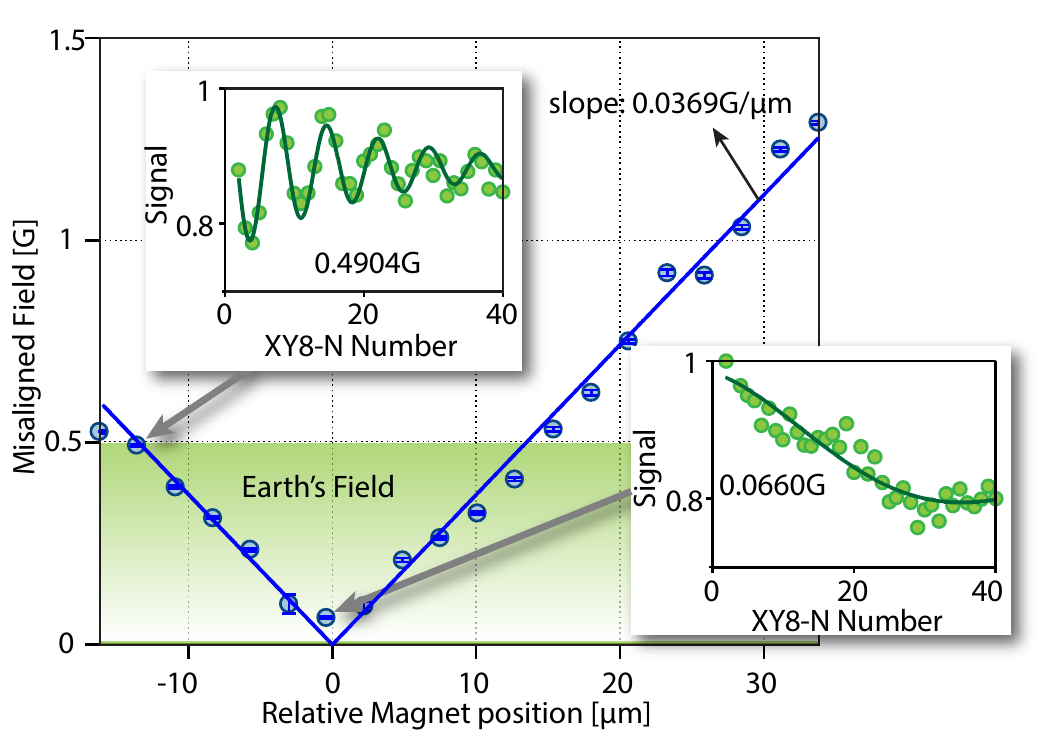}}
  \caption{\textbf{Control  of intrinsic misalignment.} Transverse field introduced by translating the magnet along the X direction at $\Delta=153$MHz. The field variation with magnet displacement is very close to linear (solid lines are a fit with slope 0.0369 G/$\mu$m), allowing one to adjust or minimize the field misalignment. The misaligned field here is estimated by XY8 experiments sweeping the cycle number (insets), with error bars obtained from a Lorentzian fit of their Fourier transforms. The shaded region represents misalignment at the $<0.5$G level, comparable to the Earth field.}
\label{fig:magnet-scanning}
\end{figure}

 \section{Interference picture of low pass signal noise filtering}
\label{app:noise-filter}
We now consider the ability of our ancilla assisted DC magnetometry protocol to effectively act as a low pass filter for the noise carried by the field to be sensed (signal noise). This signal noise could be for instance, amplitude modulations in the field originating at the signal source. Here we derive the effective filter function of the signal noise employing a first order Magnus expansion~\cite{Magnus54}. 

To derive the noise filter function, we decompose the signal noise into low frequency AC tones of frequency $f$ and obtain the signal as a function of $f$. Considering $L$ cycles of the CPMG experiment and assuming that one works in the regime, $1/T_2<f<\omega_0/2\pi$, to a very good approximation perturbation theory can be applied piece-wise (Fig.~\ref{fig:sine-wave}{(A)}) over each of the $L$ blocks, giving rise to the effective angles $\alpha_j=\gamma_eB^{(j)}_\perp A_\perp F/(\Delta\omega_0)$ proportional to the field value $B^{(j)}_\perp $ at every mean interval. Note that while considering a $\Ns$ ancilla, $\omega_0/2\pi\app 3.87$MHz, and $1/T_2\app 1$kHz,  hence our analysis is valid for a wide range of signal noise frequencies.

Now, one can evaluate the \textit{effective} propagators $\mU^{\textrm{eff}}_{\ket{0,-1}}$ over the entire sequence by taking the product, $\mU^{\textrm{eff}}_{\ket{0,-1}} = \prod_j \mU_{\ket{0},\ket{-1}}^{(j)}$. In order to illustrate the key physics of low-pass signal noise filtering, for simplicity, we consider the case of a spin-1/2 ancilla, and neglect the contributions of terms $\propto I_x$ in the perturbation theory. While this makes no difference to the noise filters obtained, it allows a dramatic simplification of the analytic formalism. For instance, since the axis $\bn_{0_\perp }$ is the same for all of the cycles, the propagator $\mU^{\textrm{eff}}_{\ket{0}}$ is particularly simple to calculate: $\mU^{\textrm{eff}}_{\ket{0}}= \exp(-i\sum_j\alpha_j\bs\cdot\bn_{0_\perp })$. 
It is evident that for high  frequency signal noise components, for which $\sum_j\alpha_j\approx 0$,  the signal is suppressed as $\mU^{\textrm{eff}}_{\ket{0}}\approx \id$. That is, there is \textit{destructive} interference between different CPMG cycles and  the noise component is effectively filtered out. It is such destructive interference, which increases proportionally to the number $L$ of cycles, that determines the signal noise filter.

By similarly evaluating $\mU^{\textrm{eff}}_{\ket{-1}}$ with a zeroth order Magnus expansion,  we obtain the signal,
\begin{widetext}
\begin{equation}
S\propto \Tr{\mU^{\textrm{eff}\dg}_{\ket{0}}\mU^{\textrm{eff}}_{\ket{-1}}} =\cos(L\sq{\mc{A}^2+\mc{B}^2})\cos(L\ov{\alpha}) - \frac{\mc{B}}{\sq{\mc{A}^2+\mc{B}^2}}\sin(L\ov{\alpha})\sin(L\sq{\mc{A}^2+\mc{B}^2}),
\end{equation}
\end{widetext}
where $\ov{\alpha} = \frac{1}{L}\sum\alpha_j$ and $
\mc{A}=-\frac{1}{L}\sum \alpha_j\sin\alpha_j\:;\: \mc{B}=\frac{1}{L}\sum \alpha_j\cos\alpha_j$. 
For weak noise, $\|\alpha_j\|\ll 1$, to first order we have  $\mc{A}\app 0\:;\: \mc{B}\app \ov{\alpha}$  and to a good approximation  $S\propto \Tr{\mU^{\textrm{eff}\dg}_{\ket{0}}\mU^{\textrm{eff}}_{\ket{-1}}} \app \cos 2L\ov{\alpha}$. Let us consider the concrete case of a single noise tone of frequency $f=1/(\qt L_p)$, where $L_p$ refers to the effective CPMG cycle number required to complete one period. Then,  assuming $\alpha_j = \alpha\cos\lb \frac{\pi j}{L_p} + \phi\rb$ and small $\alpha$, we obtain
\begin{equation}
  S\propto \cos\lcb 2\alpha\cos\lb \frac{\pi(L+1)}{2L_p} +\phi \rb \frac{\sin \lb \frac{\pi L}{2L_p}\rb}{\sin \lb \frac{\pi}{2L_p}\rb} \rcb
	\zl{filterL}
	\end{equation}
Here $\phi$  refers to the phase of the noise tone with respect to the start of the pulse sequence, and to derive the effective filter one has to perform an average over $\phi$. 
Importantly, the filter characteristics is set by the second term in the argument of \zr{filterL} which has a form analogous to a Bragg grating~\cite{Ajoy11,Ajoy13l}, with  filter width $f=1/(2L_p\qt)$ set by its first zero, and  filter suppression increasing linearly with $L$, giving a low-pass bandwidth of $1/T_2$ in the extreme case.

\begin{figure}
  \centering
  {\includegraphics[width=0.45\textwidth]{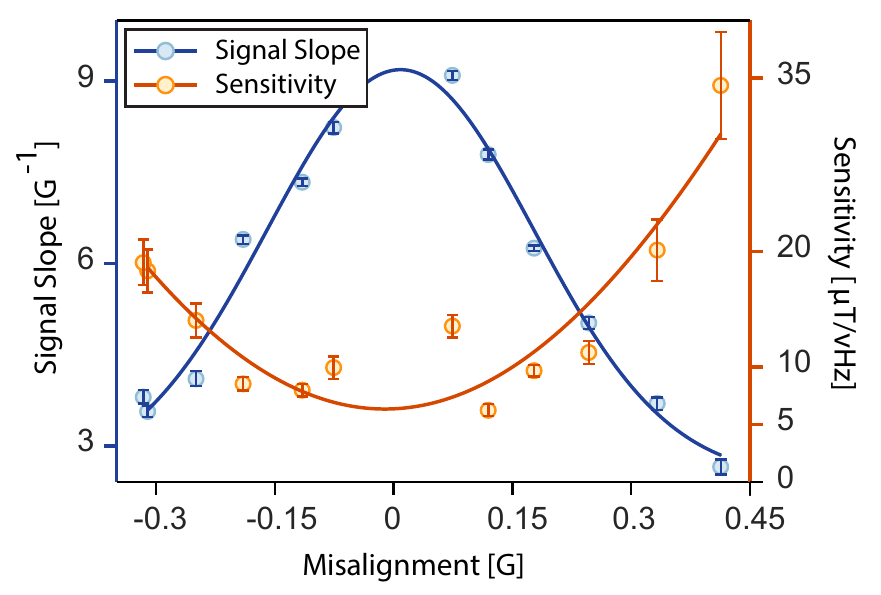}}
  \caption{\textbf{Effect of intrinsic misalignment on sensitivity.} NV magnetometer DC sensitivity (orange) and the signal slope $\left.dS/dB_e\right|_{B_0}$ excluding signal decay (blue), measured for different values of intrinsic misalignment $B_{i}$. Experiments correspond to XY8-48 (interrogation time 46.47$\mu$s) at $\Delta=174.5$MHz. Misalignment is controllably added following Fig.~\ref{fig:magnet-scanning}. Error bars are calculated from Lorentzian fits of the Fourier transforms of the obtained magnetometer fringes. The solid lines are gaussian fits to guide the eye. As is evident, increasing intrinsic misalignment has a deleterious effect on sensitivity, and for optimal sensor performance one would operate the NV sensor in the linear regime $B_e\gg B_i$, or employ magnetic field shielding. }
  \label{fig:increasing-misalignment}
\end{figure}

The action of the frequency tone has a  very clear geometric description in the space of unitary vectors (see Fig.~\ref{fig:sine-wave}{(B)}): for the propagator $\mU_{\ket{-1}}$ we have an effective shortening of the corresponding vector, while for $\mU_{\ket{0}}$ we have a fan of vectors over which the effective propagator has to be evaluated. In essence, for purely DC fields the signal terms in each cycle are all fixed strength, leading to a constructive rotation by the same amount. On the other hand, AC fields lead to scrambling of the rotations and effectively destructive interference, with the filter suppression increasing with the effective time of the sequence. Hence, our ancilla-assisted protocol leads effectively to a low pass filter for signal noise and a band-pass filter for sensor noise, both with bandwidth $1/T_2$.

In comparison,  the Ramsey method achieves poor filtering~\cite{Cywinski08,Ajoy11}, the filter being of the form $F=\lsb\sin(\pi\omega T_2^*)/\omega\rsb^2$, having bandwidth $1/T_2^*$, which could be two orders of magnitude larger than achievable in our experiment (Fig.~\ref{fig:filter-expt}). In effect then, the noise on the sensor due to spins in the NV environment set a bound on the amount of filtering one can achieve for the noise carried by the field being sensed.

\begin{figure*}[t]
  \centering
  {\includegraphics[width=0.95\textwidth]{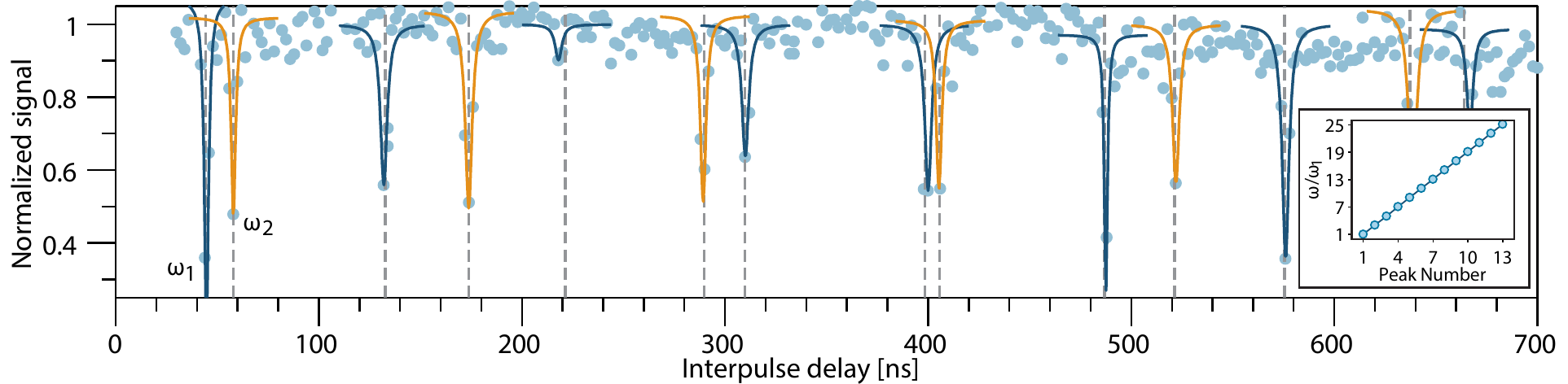}}
  \caption{\textbf{Signal harmonics in $\Ns$ assisted magnetometry.} \ DC magnetometry signal at 1344G (893 MHz) for long inter-pulse times showing families of harmonics from the $\Ns$ ancilla, occurring at $\omega_{1,2} = Q \pm A_{\pll}/2 \pm \gamma_n B_z \mp A_\perp ^2/4\Delta$. 
The two sets of harmonics, indicated by the orange and blue peaks, respectively, correspond to  nuclear spin transitions in the $m_I=\pm1$ manifolds. Gray lines indicate the theoretically expected values of the harmonics. In the inset:  frequency of the blue peaks as a function of $\omega_1$,  showing that we measure odd $\Ns$ harmonics as expected. To minimize pulse errors, one can  select an harmonic with a suitable duty cycle.} 
\label{fig:harmonic-plot}
\end{figure*}

\begin{figure}[b]
  \centering
  {\includegraphics[width=0.5\textwidth]{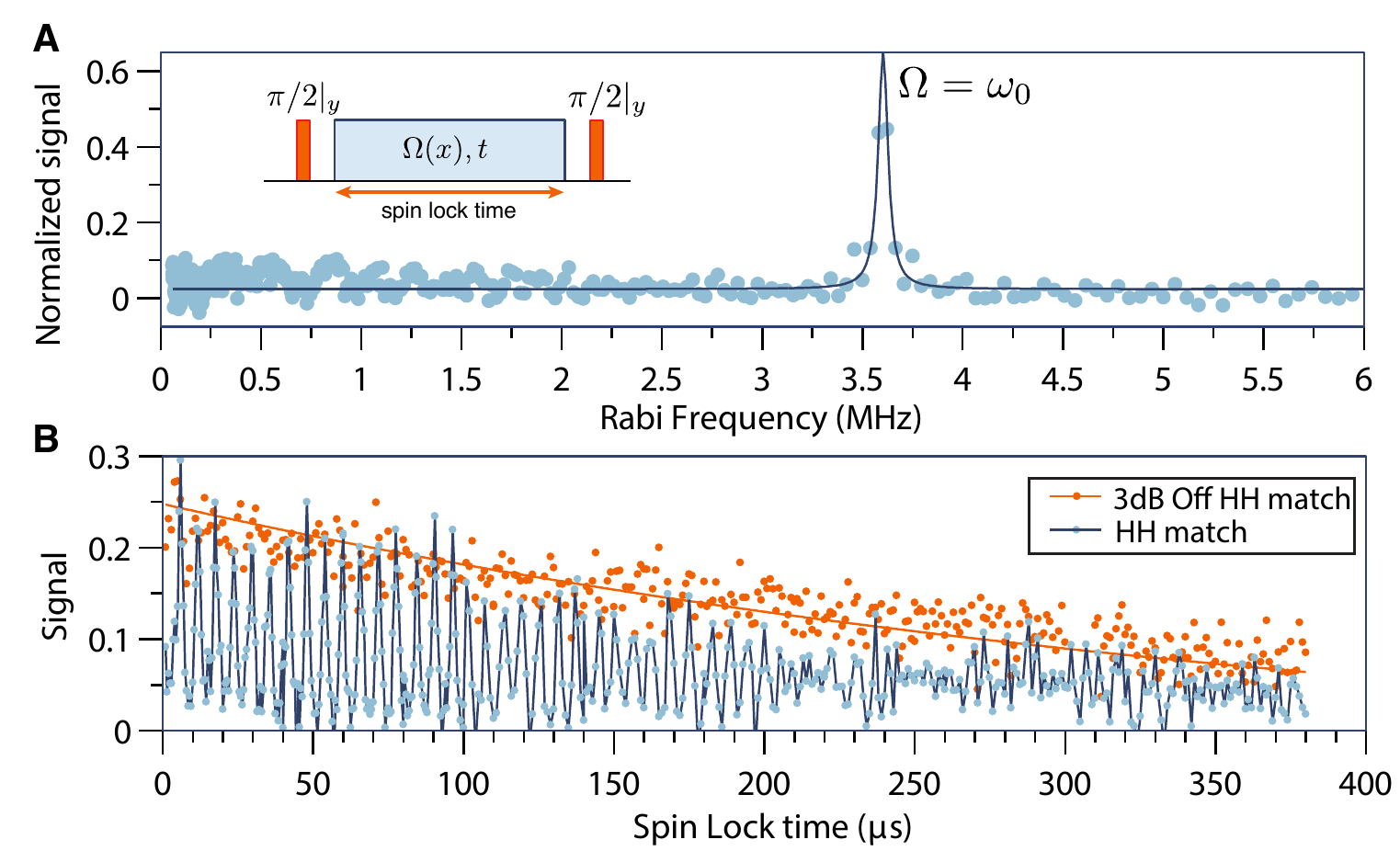}}
  \caption{\textbf{Spin lock ancilla assisted DC magnetometry} {(A)} Spin lock measurements performed at 717G, with lock length fixed at $ 10\mu$s, and sweeping Rabi frequency $\Omega$. We obtain a magnetometry signal at the Hartmann-Hahn match $\Omega=\omega_0$. Inset shows the pulse sequence. {(B)}  Sweeping spin lock time at $\Delta=165$MHz at the Hartmann-Hahn match (blue) and 3dB away (orange). The oscillations constitute the DC magnetometery signal, 
	exactly analogous to Fig.~\ref{fig:sweeping-L}. Crucially we are able to extend the sensor interrogation time $>350\mu$s.}
\label{fig:HH}
\end{figure}

\begin{figure*}[t]
 \centering
  {\includegraphics[width=\textwidth]{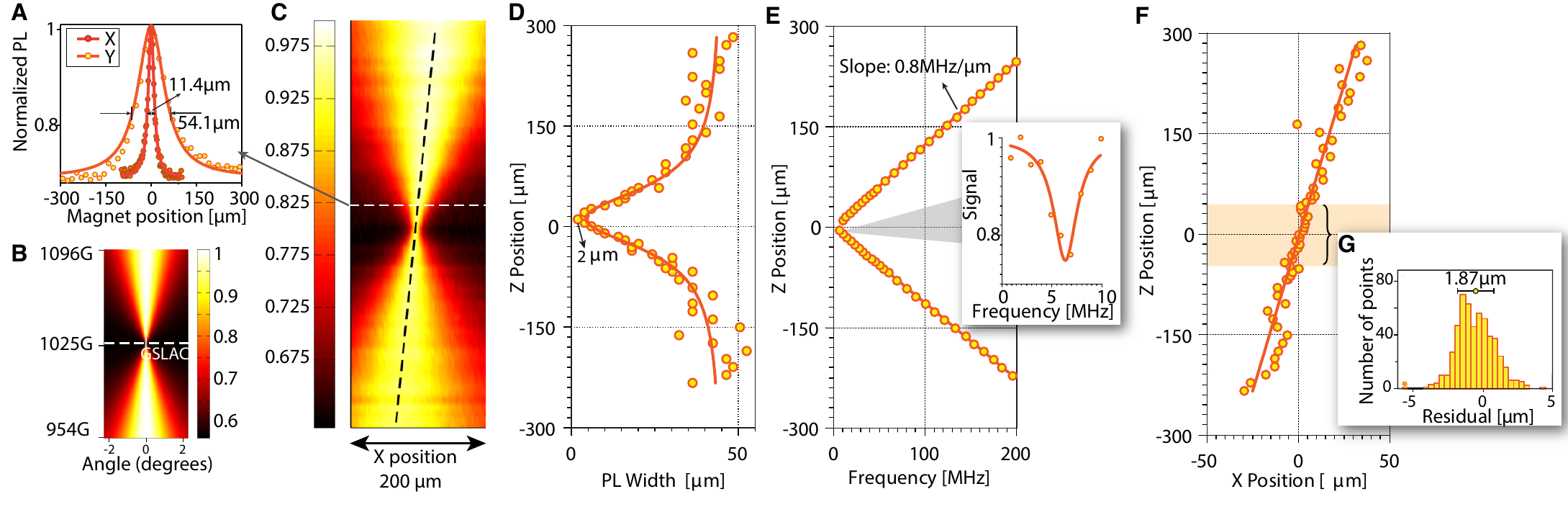}}
  \caption{\textbf{Alignment of magnetic field close to GSLAC.} {(A)}  NV center photoluminescence (PL) intensity as a function of magnet position along X and Y, showing a high sensitivity to misalignment close to the GSLAC ($B_z= 1025$G). {(B)} Simulations of NV PL as a function of misalignment angle and magnetic field. {(C)} Experimental  NV PL in an XZ magnet scan \cite{SOM}. The narrow neck is the anti-crossing point (GSLAC). The dashed black line indicates  where the PL is maximized (best alignment with the NV axis). 
 {(D)} For each Z position, we extract from a Lorentzian fit the width of the PL curves (see panel A) as a function of the magnet position X.  
The solid line is a Lorentzian fit to these widths.  At the GSLAC, the PL width is under $5\mu$m.  
{(E)}  NV electronic resonance frequency $\Delta$ as a function of the Z magnet position. The frequency varies linearly with Z (the solid line is a linear fit). We can reduce the NV resonance frequency down to $\Delta\approx 6$MHz. Inset: shows this through pulsed ODMR. {(F)} Optimization of field alignment via a gradient ascent algorithm. At each Z position, the circles are alignment positions for magnet X position. Near the GSLAC neck the PL has a larger gradient so the algorithm is more robust to noise and we can align the field better. The solid line is a linear fit only using points near the GSLAC neck (shaded). {(G)} Inset represents a histogram of the residuals to the fit in the shaded region over 36 such datasets, showing that the average error of magnet alignment close to the GSLAC is reproducibly under 1$\mu$m (about 0.1 Gauss). }
\label{fig:magnet-alignment}
\end{figure*}

\section{Effect of intrinsic misalignment on sensitivity and operating range}
\label{app:intrinsic}

We refer to intrinsic misalignment $B_{i}$ as the residual field away from the ground state anti-crossing, due chiefly to the magnetic field of the Earth. 
In principle this can be eliminated to about a part per million through magnetic shielding~\cite{Kornack07}, a conventional choice for ultra-sensitive magnetometers. However, in our experiments without any shielding,  intrinsic misalignment imposes limitations of sensor operating range, effectively setting an optimal bias point for sensor operation as we shall describe below. 

More precisely, the field to be measured $\vec B_e$, and $\vec B_i$ are both perpendicular to the N-V axis but are separated by an angle $\beta$; hence the field one measures is the vector sum $\vv{B}_\perp =\vec{B}_{e} + \vec{B}_{i}$, with $B_{\perp}= \sq{B_e^2+B_i^2 + 2B_eB_i\cos\beta}$. 
An increasing $B_e$, in general, leads to  a nonlinear response of the sensor (see Fig.~\ref{fig:parabola}). 
Indeed in the regime where $B_e\ll B_{i}$, the effective field is approximately given by $B_{\perp} \approx B_i + B_{e}\cos\beta$, leading to an effective loss of sensitivity by $\cos\beta$. 
As $\beta\to \pi/2$, the field sensitivity vanishes to first order, as increasing $B_e$ goes into changing the direction of $B_{\perp}$ rather than its amplitude. 
In the opposite regime when $B_e\gg B_i$, which we shall refer to as the \textit{linear regime}, we effectively have $B_{\perp} \approx B_i\cos\beta + B_{e}$ , for which there is no loss of sensitivity. More generally, we have that the rate of change, $dB_{\perp}/dB_e = \lsb \frac{B_{e}+B_i\cos\beta}{B_{\perp}}\rsb = \cos\xph$, where $\xph$ is the angle of the resultant vector $\vv{B}_{\perp}$ to $\vv{B}_e$ (Fig.~\ref{fig:parabola}). This allows us to concretely quantify the onset of the linear regime as being where $\xph\lesssim \pi/8$. The presence of the linear regime is evident in the experiments of Fig.~\ref{fig:sweeping-L}{(B)}, where the signal slope upon sweeping XY8-N cycle number immediately reflects $B_{\perp}$. In the figure, the solid lines are fits to the theoretical result above. Note that the values $\beta$ differ for different $\Delta$ values due to our magnet alignment procedure (\zar{magnet}). 

 In order to experimentally study the effect of intrinsic misalignment on sensitivity, we increase $B_i$ in a controllable fashion following Fig.~\ref{fig:magnet-scanning}, and perform ancilla assisted DC magnetometry at every point. The results in Fig.~\ref{fig:increasing-misalignment} indicate that intrinsic misalignment leads to a loss in sensitivity, and severely affects sensor performance; thus in experiments to characterize and operate the NV magnetometer (for instance Fig.~\ref{fig:sweeping-V}), we always work in the linear regime --  by suitable alignment of the magnetic field we can reliably reduce $B_i<0.3$G (\zar{magnet}).

\section{Robustness of $\Ns$ ancilla to magnetic field drift}
\label{app:nitrogen}
Not only has the $\Ns$ ancilla the inherent advantage of being a part of every NV center sensor, it also offers significant additional benefits from a noise rejection perspective. First, note that since the resonance frequency $\omega_0= Q-{A_{\parallel}\over 2}+\frac{A_{\perp}^2}{Q-A_{\parallel}+\Delta}$is dominated by the large quadrupolar interaction $Q_0=$-4.95 MHz ($Q=Q_0+\gamma_nB_z$), the  upconversion frequency   is much higher than for the intrinsic Zeeman field $\gamma_n B_z\app$-300 kHz alone. This not only helps in suppressing sensor noise, but it also allows a significant suppression of the signal noise. 
More remarkably, the frequency $\omega_0$ is to first order immune to magnetic field drift, since its relative change is extremely small, $\xd\omega_0/\omega_0 \app \gamma_n\xd B_z/(Q-A_{\pll}/2) = (7.75\zt 10^{-5})\cdot\xd B_z$. This ensures that the  frequency employed for quantum lock-in detection, (dynamical decoupling sequences with  quantum interpolation~\cite{SOM}), is robust against field drift. 
Even if magnetic field drifts  change  $\Delta$, this only induces  an effective amplitude modulation of the up-converted field (which is $\propto A_\perp/\Delta$), which is exactly the signal noise that can be suppressed to better than $1/T_2$ by our protocol (Sec.~\ref{sec:noise}). Hence the $\Ns$ ancilla affords an inherent robustness to magnetic field drift, and provides an enabling experimental advantage.


In order to demonstrate this experimentally (Fig.~\ref{fig:harmonic-plot}), we performed experiments at $\Delta=-893$MHz (1344G), where the $\Ns$  frequency, $\omega_0\app 3.87$MHz, is very close to what obtained  close to the GSLAC (see Fig.~\ref{fig:xy8-time}). Close to the GSLAC  we  detect only one set of echo modulation~\cite{Flanagan87,Weis98} (at $\omega_2$ in Fig.~\ref{fig:harmonic-plot}) since the $\Ns$ polarized by  optically mediated polarization exchange with the NV electronic spin~\cite{Jacques09}. However at large $\Delta$, as in Fig.~\ref{fig:harmonic-plot}, the $\Ns$ is not polarized, and the two sets of modulations at the odd harmonics of the frequencies $\omega_{1,2} = Q \pm A_{\parallel}/2 \pm \gamma_n B_z \mp A_\perp ^2/4\Delta$ are clearly discernible, which allows more choices for the up-conversion frequency $\omega_0$. The presence of the higher harmonics also points to the ability to perform experiments in a suitable pulse duty cycle regime where the pulse error can be minimized.

\section{Spin-lock DC magnetometry}
\label{app:HH}
We now consider an alternative strategy for ancilla-assisted DC magnetometry via spin locking~\cite{London13,Belthangady13} (Fig.~\ref{fig:HH}{(A)}). The NV center is spin locked at a Rabi frequency $\Omega$, and at the Hartmann-Hahn matching condition~\cite{Hartman62,Pines73} $\Omega=\omega_0$, there is once again a second-order hyperfine mixing $\propto B_\perp A_\perp /\Delta$ that causes a flow of polarization away from the NV center, which can be detected. 
In essence, this method is a rotating frame analogue of the dynamical decoupling protocol we employed in the main paper. The signal has exactly the same form of Sec.~\ref{sec:protocol}, $S=\frac{1}{2}[1+\cos(\alpha t)]$, where $t$ is the spin-lock time, and $\alpha=\gamma_eB_\perp A_\perp F/\Delta$. 
This method also benefits from  the same ancilla-assisted frequency upconversion that enabled signal and sensor noise suppression. Overall, spin-lock DC magnetometry might offer some additional  advantages: (i) the Hartmann-Hahn matching requires careful adjustment of the drive amplitude, but  does not require quantum interpolation; (ii) since decoupling is applied continuously, a larger amount of the sensor noise spectrum can be effectively suppressed, quantified by the coherence time $T_{1\xr}$ that is usually greater than echo-based $T_2$. In practice however, amplifier noise sets the ultimate coherence time achievable~\cite{Fedder11}. In Fig.~\ref{fig:HH}{(B)} we present example spin lock DC magnetometry data at $\Delta=165$MHz, indicating that an interrogation time of over 350$\mu$s is achievable. A detailed analysis of magnetometer sensitivity using spin-lock techniques will be presented elsewhere. The good performance of spin-lock DC magnetometry  indicates that  the pulsed scheme interrogation time (and hence sensitivities)  is primarily limited by pulse error, and can be mitigated through higher order pulse compensation~\cite{Souza11}. 

\section{Magnetic field alignment at the GSLAC}
\label{app:magnet}

To align the magnetic field~\cite{SOM} near the GSLAC (1025G), we exploit the fact that the NV center photoluminescence (PL) gets quenched with increasing misalignment due to level mixing in the excited state~\cite{tetienne12,van15} (Fig.~\ref{fig:magnet-alignment}{(B)}). 
We employ a permanent magnet in our experiments, and translating the magnet in the X and Y  directions to  find the position of maximum photoluminescence leads to an alignment to better than 1$^{\circ}$~\cite{SOM}. 
An experimental 2D XZ map of NV photoluminescence illustrates this more clearly, (Fig.~\ref{fig:magnet-alignment}{(C)}), the narrow neck corresponding to the GSLAC. We note that this sharp feature in  the NV photoluminescence as a function of misalignment at the GSLAC could be used as the basis for an all-optical magnetometer,  as it  was recently demonstrated~\cite{Wickenbrock16}.

The NV center resonance frequency $\Delta$ changes linearly as a function of the magnet height, and can be made to approach zero close to the GSLAC (Fig.~\ref{fig:magnet-alignment}{(D)}). 
To further optimize the alignment, down  to the intrinsic contribution given by the Earth's field ($<0.3$G), we employ a gradient ascent algorithm  to find the highest photoluminescence in the GSLAC  region (shaded), and extrapolate linearly  from this region,  exploiting  the linear behavior of the NV resonance frequency with magnet distance. The small PL variance in the GSLAC neck ensures that the alignment procedure is reproducible and accurate (Fig.~\ref{fig:magnet-alignment}{(E)}). We note that when the field is aligned with the NV axis, the $\Ns$ nuclear spin is completely polarized via a hyperfine-mediated optical process~\cite{Jacques09}, further increasing the signal contrast obtained in the magnetometry experiments. 

The amount of misalignment can be accurately measured through an XY8 experiment, sweeping the number of pulses and extracting the frequency $f_0$ of the resulting oscillations (Fig.~\ref{fig:magnet-scanning}), following $B_\perp  = \tan(f_0/2)\frac{\Delta\omega_0}{\gamma_e A_\perp F}$. We are able to reproducibly align the field to better than 0.1G, this $\sim$ 0.01\% residual error largely being set by the magnetic field of the Earth. Translating the magnet along the X axis also allows the controllable addition of misalignment in an approximately linear manner,  and enables  quantifying the effect of the residual Earth's field on the sensor dynamic range (\zar{intrinsic}).

\end{appendices}


%

\pagebreak
\newpage

\cleardoublepage

\newgeometry{margin=0.35in}

\setcounter{section}{0}
\setcounter{equation}{0}
\setcounter{figure}{0}
\setcounter{table}{0}
\setcounter{page}{1}
\makeatletter
\renewcommand{\theequation}{S\arabic{equation}}
\renewcommand{\thefigure}{S\arabic{figure}}
\renewcommand{\thetable}{S\arabic{table}}
\renewcommand{\thepage}{\roman{page}}
\renewcommand{\bibnumfmt}[1]{[S#1]}
\renewcommand{\citenumfont}[1]{S#1}

\pagenumbering{arabic}
\onecolumngrid
\begin{center}
\textbf{\large{\I{Supplementary Information:} DC Magnetometry at the $T_2$ Limit}}\\
\hfill \break
A. Ajoy, Y. X. Liu and P. Cappellaro\\
\smallskip
\emph{{\small Research Laboratory of Electronics and Department of Nuclear Science \& Engineering, Massachusetts Institute of Technology, Cambridge, MA}}\newline
\end{center}
\twocolumngrid

\beginsupplement

\section*{Summary}
In this supplementary information, we provide detailed information about the experimental setup and the setup for magnetic field alignment. Also discussed are methods sampling the signal peak using quantum interpolation, and procedures used in fitting the experimental data.

\begin{figure}[t]
  \centering
  {\includegraphics[width=0.48\textwidth]{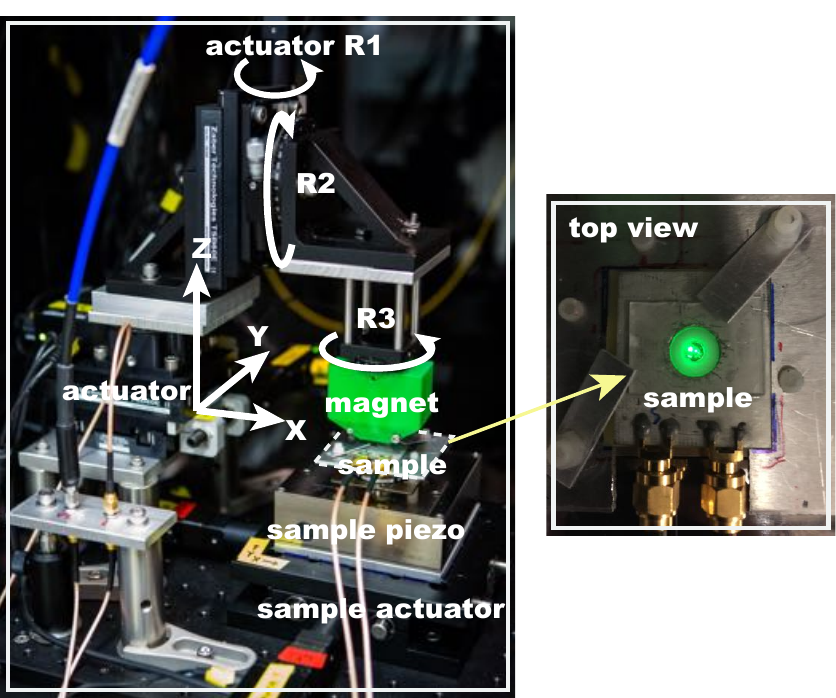}}
  \caption{\textbf{Magnet setup.} Figure illustrates the magnetic field alignment setup used in our experiments. There are four motorized actuators for linear translation of the magnet in the X, Y and Z directions, along with a rotation stage R1. R2 and R3 are two unmotorized rotation stages. The magnet is a composite assembly of 12 magnets contained in the plastic green holder. By suitable alignment protocols, the magnetic field can be aligned to better than the Earth's field at the GSLAC.}
\zfl{magnet-setup}
\end{figure}

\section{Experimental Setup}

Our experimental sample consists of isotopically pure diamond ($99.99 \%$ C-12, E6) containing NV centers produced via implantation and subsequent annealing. We used a home-built confocal microscope to address single NV centers. The details of the optical setup is identical to the one in \cite{Ajoy16x}. A 2mW 532 nm laser (SPROUT from Lighthouse Photonics) beam is first sent through an acousto-optic modulator (AOM, Isomet Corporation, M113-aQ80L-H) for switching and then focused using an oil immersion objective (Thorlabs N100X- PFO Nikon Plan Flour 1.3NA). The sample is mounted on a 3D-piezo scanner (NPoint) to scan the diamond for NV centers. The fluorescence excitation light is collected by the same objective, collimated, filtered from the 532 nm beam using a dichroic (Chroma NC338988) and then focused onto a pinhole for spatial filtering, and collected using a single-photon counting module (Perkin Elmer SPCM-AQRH-14). 

We generate microwave pulses for XY8 sequences by direct synthesis of the pulses using 1.25 GS/s four channel arbitrary waveform generator~\cite{Tabor} (Model WX1284C, Tabor Electronics Ltd.) with timing resolution of $\xD\qt=1$ns. The MW pulses are subsequently amplified using a high power amplifier (Minicircuits LZY22+) and delivered to the NV center through a 25$\mu$m Cu wire (Alfa Aeasr) terminated in a 50$\xO$ load. The AWG, the AOM and the single-photon counting module were gated using TTL pulses produced by a 500 MHz PulseBlaster (SpinCore). The DC magnetic fields using in the evaluation of sensitivity are created through voltage source (Berkeley Nucleonics BNC 575) driven on the same wire and combined with a bias tee (Minicircuits Z3BT-2R15G+). The experiments are done in a pulsed manner with a pulse rate about 0.5Hz, which ensures we operate within the bandwidth of the bias tee, and importantly also allows a concurrent measurement of intrinsic misalignment in alternate cycles to monitor thermal magnet drift. The applied voltage $V$ produces a proportional magnetic field used to characterize the magnetometer, $B_e=\xg_v V$, where from the experiments in Fig. \ref{fig:sweeping-L} of the main paper we extract the value of $\xg_v =$ 0.0407$\pm$ 0.0006 G/V. This value also is in good agreement with an independent measurement of $\xg_v =$ 0.0470 G/V from the measured NV center Rabi frequency at 91MHz, since a 0dBm RF drive produced a Rabi frequency of 10MHz. We ascribe the slight discrepancy to transmission loss of the wire at RF frequencies.

 For the experiments to determine the low pass filter function to signal noise (Fig. \ref{fig:filter-expt} of the main paper), we drive a single tone of low frequency voltage from a Rigol 1000D Arbitrary function generator through the same wire, and experimentally map out the filter function. The signal normalization in Fig. \ref{fig:filter-expt} is carried out by concurrently measuring the magnetometer signal under the corresponding RMS DC voltage and zero voltage (measuring the intrinsic misalignment).

\section{Setup for field alignment at GSLAC}

\begin{figure}[t]
  \centering
  {\includegraphics[width=0.45\textwidth]{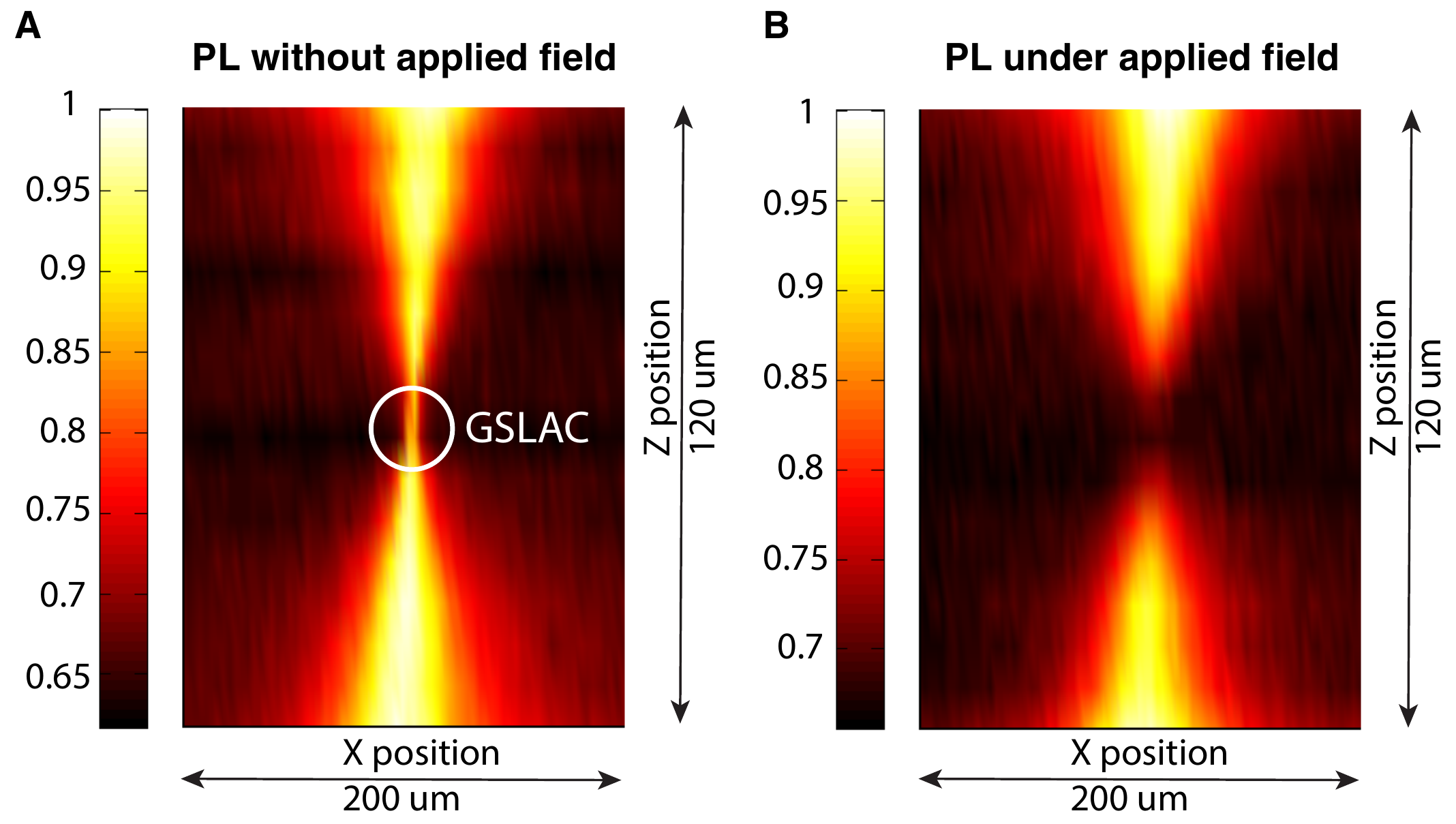}}
  \caption{\textbf{Photoluminescence quenching under applied transverse fields.} \T{(A)} On the left panel is the NV center photoluminescence (PL) obtained in an XZ magnet scan, were the thin neck region corresponds to the GSLAC (see Fig. \ref{fig:magnet-alignment} of the main paper). \T{(B)} Corresponding PL in the presence of an applied voltage of +3.6V that generates a transverse magnetic field. As is evident there is a quenching of the PL close to the GSLAC. This could form the basis for all optical vector magnetometry with a single NV center.}
\zfl{mag-PL-under-voltage}
\end{figure}

Here we describe our magnetic field alignment setup used in our experiments (see \zfr{magnet-setup}). Our magnet is a composite assembly of 10 N52 cubic magnets of 9.53mm edge (K\&J Magnetics B666-N52) and 1.48T surface magnetization. The magnets are held together in a plastic case that was 3D printed. The magnet design was based on the construction used by the Ulm group~\cite{Unden16}.  We define X to be the effective magnetization axis (N-S) of the magnet assembly, and is placed parallel to the edge of the [100] diamond. The magnet is a mounted  (\zfr{magnet-setup}) on a composite XYZ translation and rotation stage controlled by motorized actuators (Zaber TNA08A50 and Zaber RSW60A-T3 respectively). There are two more manual rotor stages for fine adjustment to ensure that the magnet surface is parallel to the sample edge. Magnet alignment close to the GSLAC is carried out primarily with the actuator based linear translation stages, details of which are presented in Appendix \ref{app:magnet} of the main paper. The motorized rotation stage R1 (\zfr{magnet-setup}) allows a rapid transfer from the GSLAC to close to zero-field, and is used for debugging purposes.

NV center photoluminescence in a typical XZ magnet scan close to the GSLAC is shown in \zfr{mag-PL-under-voltage}\T{(A)}, where the narrow neck indicates the high sensitivity to field misalignment at the GSLAC. In the presence of a strong transverse field, applied through an external voltage, there is quenching of the photoluminescence, that can be exploited for all-optical vector magnetometry~\cite{Wickenbrock16x}.

\section{Quantum interpolation to efficiently sample the signal peak}
In this section we describe the method of ``quantum interpolation" that was employed in our experiments in order to precisely lock into the up-converted $\Ns$ signal. The motivation for using quantum interpolation is that the delay between the $\pi$ pulses in XY8 experiments has to be $\pi/\omega_0$, however this is often limited in precision in the finite timing resolution $\xD\qt$ of the hardware used to generate the pulses (in our case $\xD\qt=1$ns). Indeed as the number of pulses $N$ increases, there is a significant loss of contrast if one does efficiently sample the true signal peak. The error goes as~\cite{Ajoy16x},
$
\xe \app \fr{1}{4}(N\xa)^2\xD\qt^2\lb 2 - \fr{\xa^2}{2}\rb^2
$
that scales $\propto N^2(\xD\qt)^2$, leading consequently to a severe loss of magnetometer sensitivity at large $N$.

\begin{figure}[t]
  \centering
  {\includegraphics[width=0.5\textwidth]{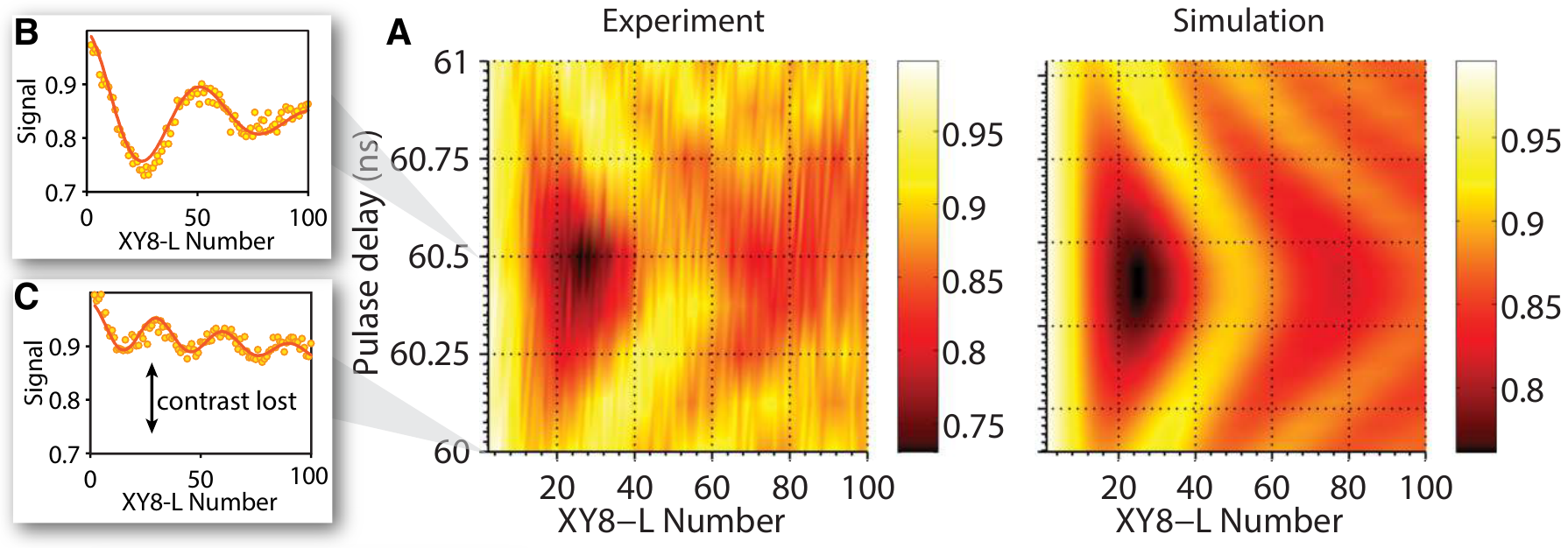}}
  \caption{\textbf{Quantum interpolation to efficiently sample the signal peak.} \T{(A)} In a $\xD\qt=1$ns window about the  $\Ns$  signal peak (Y axis), we sweep the XY8-L number (X axis), with $\xD\qt$ being the timing resolution of our timing hardware (see also Fig. \ref{fig:xy8-time} of the main paper). These experiments are performed at $\xD=$150MHz. We use quantum interpolation effectively supersample 16 slices in this window. The signal obtained matches closely with the expected theoretical lineshape in Sec. \ref{sec:protocol} of the main paper (right panel), allowing us to estimate the static misaligned field in these experiments $|B_\perp | = 6.83\mu$T. \T{(B-C)} Lower inset demonstrates that without quantum interpolation there is significant loss of signal contrast, but the full contrast is recovered if one samples the signal peak (upper inset).}
\zfl{nitrogen-2D}
\end{figure}

We recently developed quantum interpolation~\cite{Ajoy16x} as a means to overcome this hardware problem, allowing one to effectively \I{supersample} the signal peak with arbitrarily high precision even with a finite $\xD\qt$. If the propagator $\mathcal U(\tau_k)$ describes action of a CPMG pulse sequence block, with $\pi$-pulses separated by a time $\tau_k = k \Delta \tau$, then quantum interpolation refers to the construction,
\begin{equation}
U^N(\tau_{k+ p/N}) =\mathcal P\left\{\prod_{m=1}^{N-p}\mathcal U(\tau_k)\prod_{n=1}^{p}\mathcal U(\tau_{k+1}) \right\} \ \approx \mathcal U^N(\tau_{k+ p/N}),
\label{eq:naiveinterp}	
\end{equation}
that allows the construction of supersamples $\tau_{k+ p/N}$ that cannot usually be accessed. Here the permutation $\mathcal P$ refers to a optimal sequence ordering that minimizes error in the construction, the error going as $\mathcal O(\Delta \tau^2)$. Importantly the  number of additional timing samples achievable via quantum interpolation scales linearly with the total number of pulses~\cite{Ajoy16x}, allowing us to completely mitigate the problem of hardware finite hardware resolution for increasing $N$.  

In Fig. \ref{fig:xy8-time} of the main paper quantum interpolation is used to sample the $\Ns$ signal peak to about 48 times higher resolution (20.8ps) than set by our hardware, allowing us to to determine the exact peak position at $\app 60.45$ns. The signal lineshape
can be described as a modified sinc-function with a slight asymmetry about the peak, and can be evidently discerned in Fig. \ref{fig:xy8-time} of the main paper. The power of the technique is demonstrated in \zfr{nitrogen-2D}, where we obtain the magnetometry signals similar to Fig. \ref{fig:xy8-time} of the main paper for 16 different supersamples in the $\xD\qt$ window set by our hardware. There is significant signal loss if one does not sample the true $\Ns$ signal peak \zfr{nitrogen-2D}\T{(B-C)}. The figure also illustrates that the experimental data closely matches the theoretically expected sinusoidal lineshape. The obtained ``\I{chevron}" patterns in \zfr{nitrogen-2D}\T{(A)} are exactly identical to those in two-spin exchange~\cite{Ajoy12x}. Overall, this illustrates that the quantum interpolation construction is indeed of low error, and is critical to obtaining the true $\Ns$ signal peak used in ancilla assisted DC magnetometry.

\section{Data analysis}
For clarity, we now detail the procedure for the fitting of experimental data that is used to calculate the corresponding error bars, primarily in Fig. \ref{fig:sweeping-V}, Fig. \ref{fig:filter-expt} and Fig. \ref{fig:sweeping-L} of the main paper.  We use a Monte-Carlo approach to estimate the uncertainty of the various fit parameters following Ref.~\cite{Ajoy16x}. 

Let us denote the fit parameters for our model by $\mathbf P$. For a given set $\mathbf P$, our theoretical model provides a non-linear functional relation $y=f(x | \mathbf P)$. Given a measured set of data points $\{x_n\}$ and $\{y_n\}$, we determine the optimal set of parameters $\mathbf P_{\mbox{\tiny opt}}$ by minimizing $\chi^2 = \sum_n [y_n-f(x_n | \mathbf P)]^2/\sigma_y^2$. Here we have assumed that the statistical error $\sigma_y$ of the measured data points is identical for all points.

Once $\mathbf P_{\mbox{\tiny opt}}$, the statistical uncertainty of $y_m$ is estimated from the deviation from the optimally fitted function $\sigma_y^2 \approx {\sum_n [y_n-f(x_n| \mathbf P_{\mbox{\tiny opt}})]^2 }/{(N-1)}$, where $N$ is the number of data points. The value of $\sigma_y$ obtained by this procedure yields sets a lower bound for the true statistical uncertainty, as any systematic deviation of the fitted function (i.e. if we have not captured the underlying true functional form in our theoretical model) increases $\sigma_y$. Subsequently the uncertainty in the fit parameters $\mathbf P$ can be estimated beyond linear order by generating artificial data sets of points $\{x_n\}$ and $\{y_n\}$ statistically distributed around $f(x_n| \mathbf P_{\mbox{\tiny opt}})$, subsequently performing a fit for each data set. We assume a Gaussian distribution for the generation of these data points, an assumption which can be verified by inspecting the distribution of $\delta y_n=y_n-f(x_n| \mathbf P_{\mbox{\tiny opt}})$ in the original data. Repeating this procedure yields a distribution of fit parameters of which the distributional form, confidence intervals and standard deviation for the individual parameters can be extracted.

\end{document}